\documentclass[11pt]{article}

\usepackage{xcolor,fullpage}
\usepackage{pslatex}
\usepackage{graphicx}
\usepackage{fancyhdr}
\usepackage{amsmath}
\usepackage{amssymb}
\usepackage{textcomp}
\usepackage[square,sort,comma,numbers]{natbib}
\usepackage{subcaption}	
\usepackage{hyperref}
\usepackage{setspace}
\usepackage{cancel}
\usepackage{mathtools}
\usepackage[font=footnotesize]{caption}
\usepackage{rotating}
\usepackage{adjustbox}
\usepackage{multirow}
\usepackage{tabularx}
\usepackage{lipsum}
\usepackage{booktabs}
\usepackage{subfiles}
\usepackage{titling}
\usepackage{placeins}
\usepackage{authblk}
\usepackage{comment}
\usepackage{wrapfig}

\newcolumntype{L}[1]{>{\raggedright\let\newline\\\arraybackslash\hspace{0pt}}m{#1}}
\newcolumntype{C}[1]{>{\centering\let\newline\\\arraybackslash\hspace{0pt}}m{#1}}
\newcolumntype{R}[1]{>{\raggedleft\let\newline\\\arraybackslash\hspace{0pt}}m{#1}}

\title{Uncovering wall-shear stress dynamics from neural-network enhanced fluid flow measurements}
\author[1]{Esther Lagemann$^*$}
\author[1]{Steven L. Brunton}
\author[1]{Christian Lagemann$^*$}
\affil[1]{\small Department of Mechanical Engineering, University of Washington, Seattle, WA 98195, United States}
\affil[$*$]{{\footnotesize  contributed equally}}

\begin{document}
\date{}
\maketitle

\vspace{-2cm}

\begin{abstract}
\normalsize
\noindent
Friction drag from a turbulent fluid moving past or inside an object plays a crucial role in domains as diverse as transportation, public utility infrastructure, energy technology, and human health. As a direct measure of the shear-induced friction forces, an accurate prediction of the wall-shear stress can contribute to sustainability, conservation of resources, and carbon neutrality in civil aviation as well as enhanced medical treatment of vascular diseases and cancer. Despite such importance for our modern society, we still lack adequate experimental methods to capture the instantaneous wall-shear stress dynamics. 
In this contribution, we present a holistic approach that derives velocity and wall-shear stress fields with impressive spatial and temporal resolution from flow measurements using a deep optical flow estimator with physical knowledge. The validity and physical correctness of the derived flow quantities is demonstrated with synthetic and real-world experimental data covering a range of relevant fluid flows.

\end{abstract}

\section{Introduction}
Whenever a fluid flow passes a surface, the induced velocity gradient generates tangential stresses, known as wall-shear stress, at the fluid-structure interface. The precise knowledge of the wall-shear stress and the associated friction forces is highly relevant in various domains ranging from the transportation sector~\citep{marusic2010predictive,yang2013,sparreboom2009principles,cooper2004} and the public utility infrastructure~\citep{grant2012taking,chung2021natural,duan2020} to energy conversion~\cite{musa2018,dalili2009,chamorro2013} and human health related areas~\citep{bellien2021,zhou2017,adamo2009biomechanical, tzima2005mechanosensory,franke1984induction}. For instance, accurate wall-shear stress predictions are essential for the development of friction drag reduction techniques in civil aviation~\citep{quadrio2022,marusic2021,albers2021} with a direct impact on fuel efficiency and emissions. 
Fuel consumption scales almost linearly with the aerodynamic drag at cruise condition~\cite{Ricco2021}, so drag reduction of only a few percent can substantially reduce the required fossil fuels and the aircraft emissions. 
Access to instantaneous, highly resolved wall-shear stress information is also medically relevant, since shear-related friction forces can promote vascular dysfunction and atherosclerosis of human arteries~\citep{brown2016,hartman2021,souilhol2020}. 
Recent research has also suggested that wall-shear stress in the lymphatic vasculature plays a crucial role in cancer progression and metastasis~\citep{lee2017}.

However, these efforts are currently challenged by a critical lack of accurate spatial and temporal resolution of the wall-shear stress dynamics for real-world applications. 
Numerical simulations can provide high spatial and temporal resolutions, but the computational costs associated with high-fidelity simulations currently limit these computations to rather simple flows with limited physical complexity, e.g., low flow speeds and basic geometries~\citep{vinuesa2022}.
Experimental measurements provide access to much more realistic flow conditions, but are handicapped by a notorious lack of methods that simultaneously capture the temporal and the spatial evolution of the wall-shear stress~\citep{orlu2020}.
For instance, surface hot films~\citep{alfredsson1988,colella2003,abbassi2017} and wall-mounted hot-wire probes~\citep{sturzebecher2001,mathis2013,gubian2019} can only measure temporal fluctuations at a single location, while oil-film interferometry~\citep{tanner1976,segalini2015,lunte2020} only visualizes the spatial wall-shear stress without temporal dynamics. 
Alternatively, emerging research focuses on developing empirical models to infer the wall-shear stress from velocity data further away from the wall~\citep{marusic2010predictive,mathis2011,balasubramanian2023,maeteling2022} or from limited sensor measurements~\citep{arzani2021,arun2023}. However, such models are not yet used as a replacement for actual wall-shear stress data because they possess limited validity and generalizability, they typically rely on empirical coefficients, and there is no general agreement on a single model representative of all physical processes.

To overcome the existing constraints, we propose a novel workflow that combines experimental measurements and data-driven modeling (i.e., machine learning) to provide accurate wall-shear stress dynamics with an exceptional spatial and temporal resolution. We rely on a well-established optical velocity measurement technique called particle-image velocimetry (PIV)~\citep{raffel_particle_2018,scharnowski2020} since it is widely used in fluid dynamics laboratories. PIV derives the flow field from images of tracer particles, which are added to the flow for visualization purposes. The particle displacement between two consecutive images and their interframing time is used to calculate the velocity distribution. In traditional PIV processing, the original image pairs are divided into sub-windows, which are cross-correlated to obtain an averaged velocity estimate per window~\citep{raffel_particle_2018}. This statistical approach inherently coarsens the spatial resolution of the resulting velocity field compared to the original image size. For instance, a generic PIV evaluation of particle image pairs containing $1024 \times 1024$~px$^2$ results in just $124 \times 124$ velocity vectors. Although the wall-shear stress can theoretically be derived from the fluid velocity close to the wall by physical laws~\citep{pope2000}, the typically low data resolution and the largely underpredicted velocity gradients close to the wall~\citep{kahler2012} usually inhibit such a derivation from PIV data. Existing approaches specifically tailored to the wall-shear stress estimation like the Clauser chart method~\citep{clauser1956,fernholz1996} and the single-pixel ensemble correlation~\cite{westerweel2004,kahler2006} are restricted to time-averaged quantities, i.e., wall-shear stress dynamics cannot be uncovered.

An emerging body of research has successfully demonstrated how fluid dynamics research can benefit from incorporating machine learning techniques~\citep{brunton2020,li2020,kochkov2021,bae2022,vinuesa2023}. In this respect, deep optical flow networks~\citep{ilg2017,teed2020} were particularly tailored to enhance PIV processing. By maintaining the original image resolution in the resulting velocity field, the recently proposed deep learning framework RAFT-PIV~\citep{lagemann2021,lagemann2022} avoids the spatial resolution reduction and the gradient smoothing of established PIV routines without compromising on physical accuracy. For the example given above, RAFT-PIV yields $64$ times more velocity vectors and an eightfold resolution refinement in each spatial direction. Consequently, the combination of a deep optical flow network with physical laws constitutes a promising approach to derive accurate wall-shear stress measures with outstanding spatial and temporal resolution from PIV velocity measurements.

\begin{figure}[]
   \centering
    \includegraphics[width=0.99\textwidth]{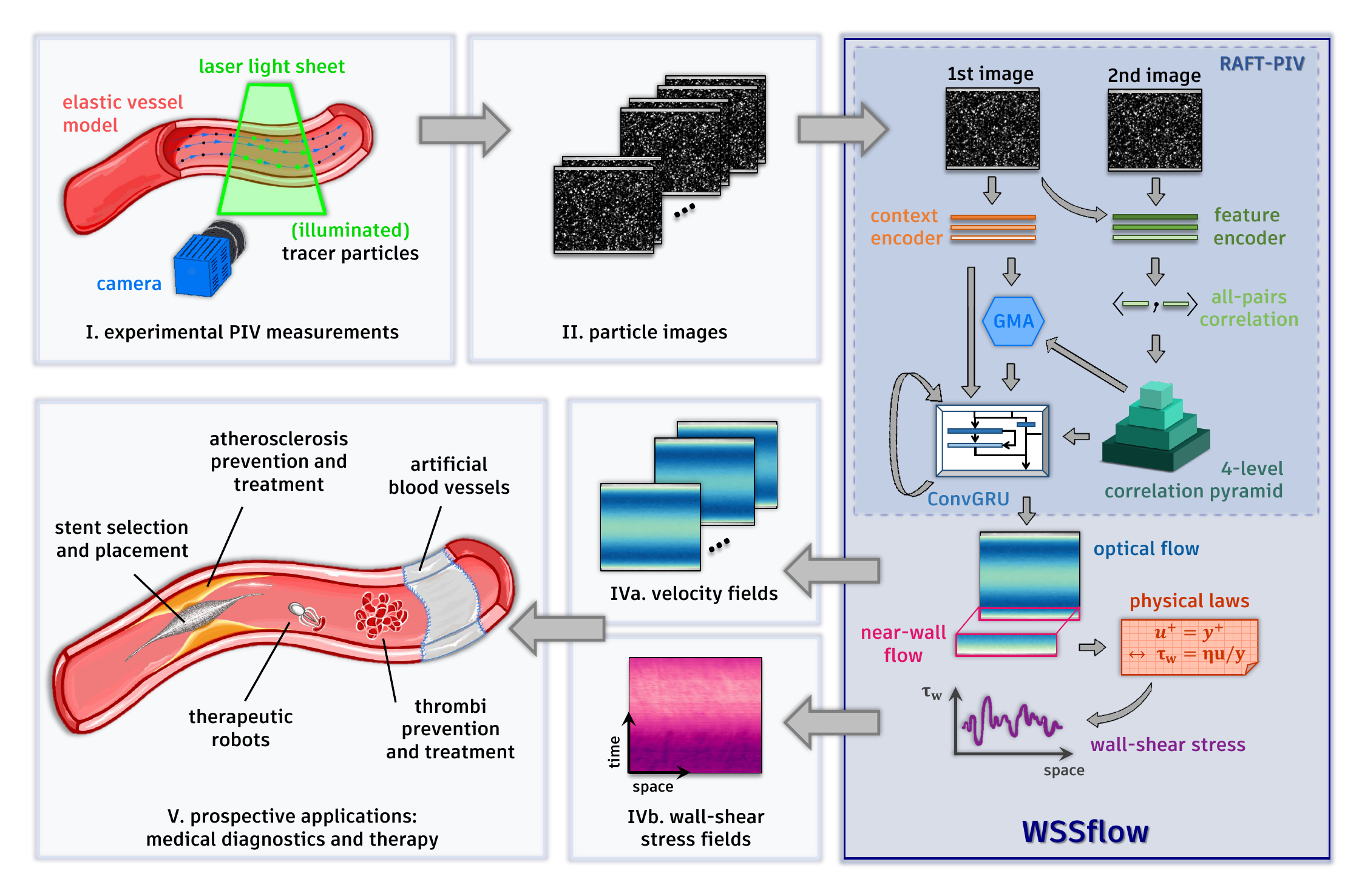}
    \caption{\textbf{Workflow using the novel deep learning framework \textit{WSSflow} to estimate velocity and wall-shear stress fields exemplified by the elastic blood vessel model.} I.)~Experimental particle-image velocimetry (PIV) measurements with the fluid dynamical model are performed. The fluid flow is observed via tracer particles, which are illuminated by a pulsed laser light sheet and captured with a high-quality camera. II.)~Gray-scale sequential particle image pairs are obtained from the PIV measurement with the bounding vessel wall at the top and bottom of each image. III.)~Each particle image pair is evaluated by the deep learning framework WSSflow, which outputs an optical flow at the input image resolution and a high resolution wall-shear stress distribution. In detail, both images are processed by a shared feature encoder, a subsequent all-pairs correlation, and a 4-level correlation pyramid. The output is fed into a global motion aggregation module (GMA) together with the context encoded first particle image. The GMA addresses image occlusions by transferring flow field information from non-occluded regions to image parts with low texture. The GMA output, the context encoded first particle image, and the correlation pyramid output are processed by a Convolutional Gated Recurrent Unit (ConvGRU), which iteratively updates the flow prediction until the final optical flow is obtained. The flow field close to the wall is extracted and combined with physical laws to derive the wall-shear stress $\tau_w(x) = \eta u(x)/y$, where $u(x)$ is the streamwise velocity at streamwise location $x$, $\eta$ the dynamic viscosity of the fluid, and $y$ the wall-normal location at which $u$ is measured. IV.)~ WSSflow provides (a) a time series of two-dimensional high-resolution velocity fields and (b) the temporal evolution of the spatially developing wall-shear stress. V.)~The obtained flow field information can be used prospectively for medical diagnostics and therapy.}
    \vspace{-0.2cm}
    \label{fig:overview}
\end{figure}
Our proposed workflow, incorporating the novel framework \textit{WSSflow}, is visualized in figure~\ref{fig:overview} based on experimental measurements of a blood vessel model. WSSflow ties in with the success of RAFT-PIV to derive the wall-shear stress dynamics from optical-flow based high-resolution velocity distributions close to the wall. More precisely, we use the fact that the flow in the viscous sublayer, which is a very thin wall-parallel layer adjacent to the boundary, is dominated by viscous forces. This allows a direct calculation of the wall-shear stress $\tau_w(x,t)$ as a function of time $t$ and streamwise location $x$ from the streamwise velocity $u(x,y,t)$, the dynamic viscosity of the fluid $\eta$, and the wall-normal location $y$ at which $u$ is measured. The only prerequisite for this approach is that the measurement setup captures the viscous sublayer by at least one pixel, which is no limiting factor in the majority of applications. 

We verify the successful estimation of time-averaged and instantaneous fluid flow quantities with fluid dynamical test cases for which either a ground truth distribution or an analytic solution exists. Apart from synthetically generated particle images, we also demonstrate a convincing quality of the wall-shear stress estimation in real experimental settings that are subject to measurement uncertainties. Besides important academic configurations like turbulent channel flows, we further demonstrate the generalization ability of our framework on challenging experimental measurements of an elastic blood vessel flow. Overall, the success of WSSflow on this broad range of fluid flows provides compelling evidence of a new capability to obtain quantities of interest, such as wall-shear stress, that were previously inaccessible with existing experimental and computational approaches. 

\section{Results}\label{sec:results}
As an effective workflow to extract temporal and spatial velocity and wall-shear stress distributions from PIV measurements, our neural processing tool WSSflow presents a significant practical advance for many real-world applications. To demonstrate the generalization ability of our method, we carry out systematic experiments in three representative measurement campaigns. The capability of WSSflow to extract true physical quantities with higher resolution is verified by comparisons with a state-of-the-art (SOTA) PIV routine and with complementary high-fidelity numerical simulations and analytic solutions, which are referred to as \textit{baseline} data. Details of the algorithms and the baseline data are provided in the Supplementary Information.

\begin{figure}[]
   \centering
    \includegraphics[width=0.99\textwidth]{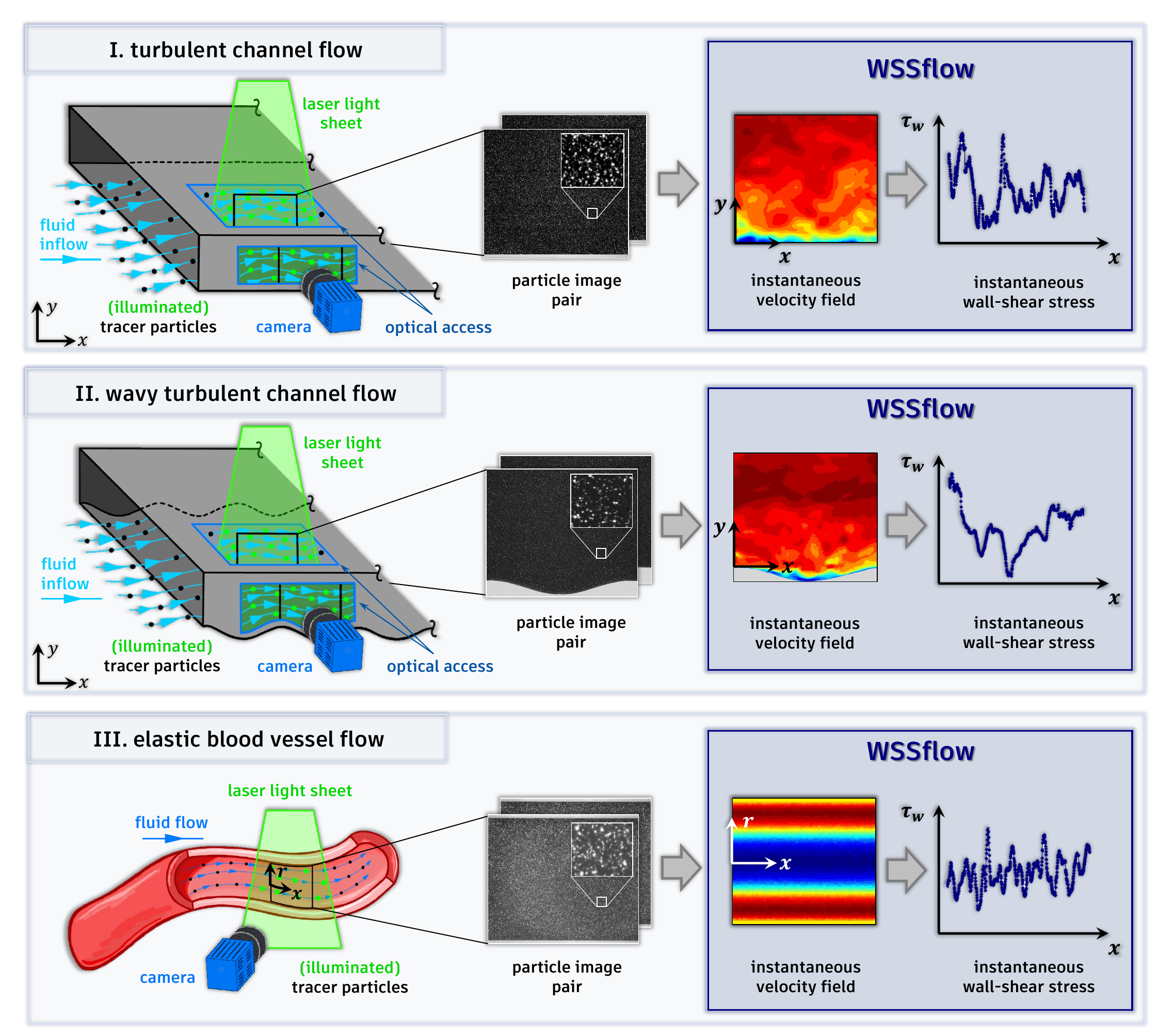}
    \caption{\textbf{Outline of the experimental measurements and data processing steps.} Three test cases are investigated in this study: I. a turbulent channel flow (top), II. a wavy turbulent channel flow (center), and III. an elastic blood vessel flow (bottom). At first, PIV measurements are conducted for each configuration. A pulsed laser light sheet illuminates tracer particles, which are added to visualize the flow, and particle image pairs are recorded with a high-speed camera. From each image pair of the time series, the novel framework WSSflow extracts the instantaneous velocity field $u = f(x,y)$ and the instantaneous spatial wall-shear stress distribution $\tau_w = f(x)$. \\\hspace{\textwidth} Both turbulent channel flows are investigated in an Eiffel-type wind tunnel with an exchangable side wall allowing to switch between a flat (case I) and a wavy (case II) lower channel wall. The sketches represent an extract of the measurement section equipped with two glass windows for optical access. The measurement plane is oriented in the streamwise ($x$) and wall-normal ($y$) direction with the fluid flow from left to right.
    The physical blood vessel model (case III) is made from transparent silicone to allow optical access. It is installed in a water tunnel facility specifically designed to study a variety of human blood vessel models. The oscillatory pipe flow is generated with a piston pump, while two reserviors generate the desired transmural pressure. A heating circuit is tempering the water-glycerin mixture to $25.6^\circ$C such that the rheological properties of the fluid replicate those of human blood at $37^\circ$C. The measurement plane is oriented in the streamwise ($x$) and radial ($r$) direction. \\\hspace{\textwidth} Please note that the sketches of the experimental setups are not to scale. Technical drawings and detailed descriptions are provided in the Supplementary Information.}
    \vspace{-0.2cm}
    \label{fig:experimental_setup}
\end{figure}

\paragraph{Datasets}\label{subsec:datasets}
We consider two academic configurations frequently used for fundamental turbulence research, namely flat-plate and wavy turbulent channel flows, and a real-world inspired oscillating flow through a flexible blood vessel model. Illustrations of the experimental setups and the data processing workflow are provided in figure~\ref{fig:experimental_setup}. 

The turbulent channel flow refers to the turbulent flow through a rectangular duct, where the length and the width of the duct are much larger than the height. An important measure to quantify the respective flow conditions is the friction Reynolds number $Re_\tau = u_\tau h/ \nu$, where $u_\tau$ is the friction velocity and $\nu = \eta/\rho$ the kinematic viscosity with $\eta$ being the dynamic viscosity and $\rho$ the fluid density. Since the friction velocity is directly related to the wall-shear stress $\tau_w = \rho u_\tau^2$, a precise knowledge of the wall-shear stress is of high significance for the flow state quantification and for fundamental turbulence research. 

The wavy turbulent channel flow is a turbulent channel flow with a sinusoidal side wall on one side of the rectangular duct. Thus, this flow exhibits changing pressure gradients along the streamwise direction creating challenging flow topologies. Characteristics like an unsteady shear layer, recirculation, and local detachment make it a representative configuration for other flows that experience similar features in a variety of real-world applications.

For both channel flows, synthetic and real-world experimental data are analyzed. Synthetic particle images are necessary to compare the WSSflow output against a known ground truth. Since the true underlying flow field of experimental PIV data is unknown on an instantaneous level, these datasets can only be compared with the baseline for time-averaged (statistically converged) quantities. However, since we are specifically interested in the wall-shear stress distribution at each individual time instant, we have to rely on synthetic data to demonstrate the accuracy of the WSSflow output. To generate realistic and physically significant synthetic particle images, we use the flow fields obtained from high-fidelity numerical simulations as the underlying flow topology. The rendering pipeline is described in the Supplementary Information. 

Bridging the academic and the real world, an experimental model of an elastic blood vessel is investigated as a third test case. Since cardiovascular diseases are the leading cause of death worldwide, a detailed understanding of how such diseases emerge and develop is of utmost importance. One key aspect is the intricate interaction of the blood flow and the elastic vessel wall since it influences the endothelial cell function and phenotype. A direct measure of this interaction is the wall-shear stress. Thus, a precise knowledge of its instantaneous and spatially developing behavior is an inevitable requirement for human health research. Here, we study the oscillating flow through a transparent elastic vessel model at flow conditions similar to the human abdominal aorta.

\subsection{Turbulent channel flow}\label{subsec:TCF}
\begin{figure}[h]
   \centering
    \includegraphics[width=0.95\textwidth]{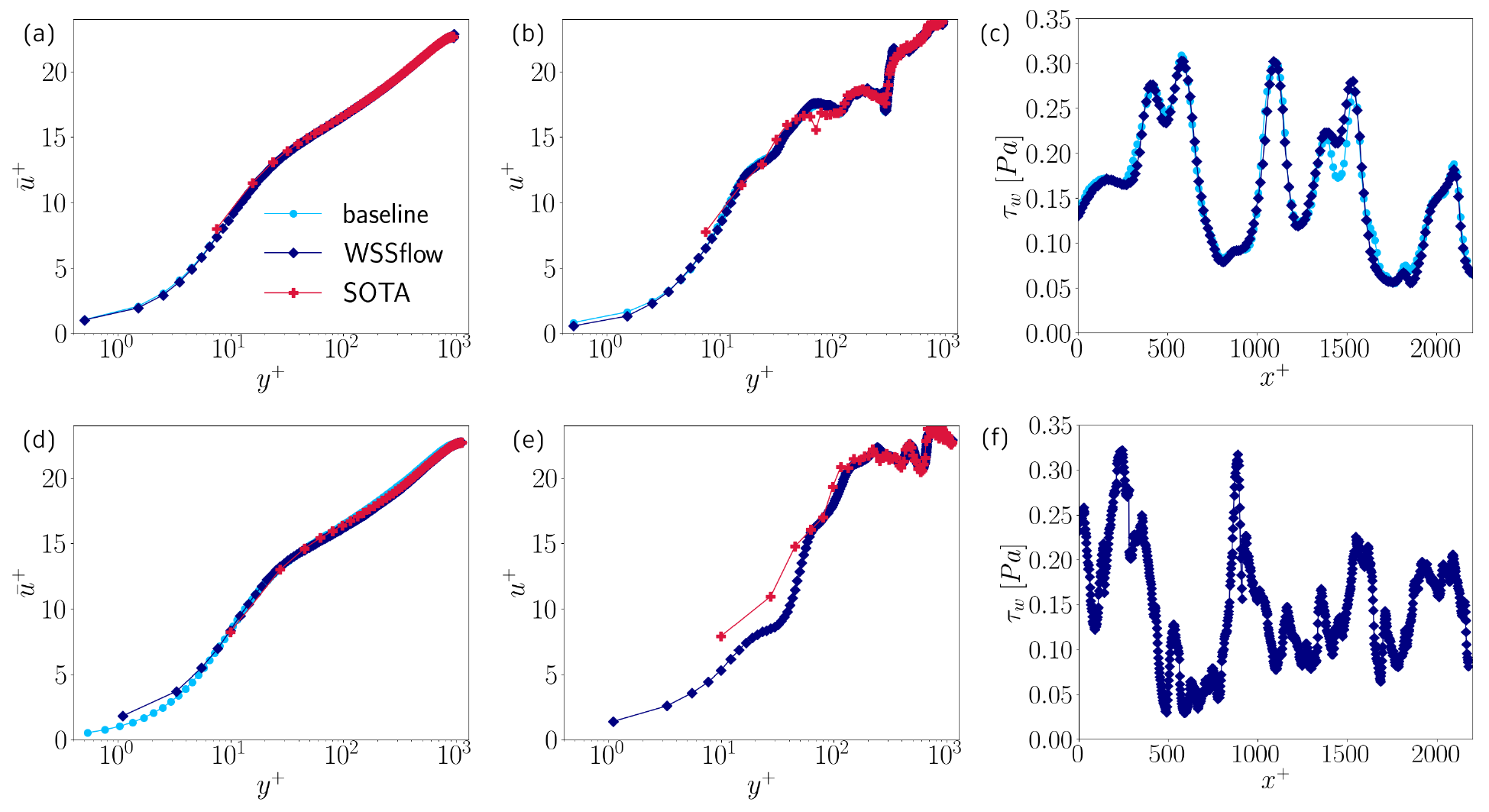}
    \caption{\textbf{Velocity and wall-shear stress distributions of the turbulent channel flow data.} Figures in the upper row (a-c) are related to the synthetic particle images and the lower row (d-f) provides findings based on the experimental data. (a,d) depict the averaged streamwise velocity profile $\bar{u}^+$ as a function of the wall-normal distance $y^+$, where the $+$ sign denotes scaling in inner units, i.e., a normalization by the friction velocity $u_\tau$ and the kinematic viscosity $\nu$. (b,e) show the instantaneous inner-scaled velocity profiles $u^+$ of an arbitrary time step and streamwise position as a function of $y^+$. (c,f) present the instantaneous wall-shear stress distributions $\tau_w$ of an arbitrary time step along the inner-scaled streamwise direction $x^+$. The velocity profiles and the instantaneous wall-shear stress distribution obtained from the synthetic data (a-c) via WSSflow follow the baseline remarkably well. The velocity profiles of the traditional PIV processing tool SOTA have a much coarser spatial resolution and do not provide velocity information in the buffer layer and below ($y^+ \lesssim 30$). Moreover, the instantaneous velocity profile deviates severely from the true underlying distribution in the lower logarithmic layer and below ($y^+ \lesssim 100$). Since the experimental data do not possess a known ground truth on an instantaneous level, only the time-averaged velocity profile can be compared to the baseline data which match extremely well. The frequency and magnitude of the experimental wall-shear stress fluctuations (f) is comparable to the synthetic data obtained from a high-fidelity numerical simulation, which demonstrates a physical significance in a qualitative sense although a quantitative comparison is not possible.}
    \label{fig:TCF}
\end{figure}

The streamwise velocity and the wall-shear stress distributions estimated by the proposed deep learning framework WSSflow are given in figure~\ref{fig:TCF}. The upper row (a-c) depicts quantities from the synthetic dataset and the lower row (d-f) is related to real-world experimental data. Where available, ground truth information is provided via the baseline data and estimated quantities based on SOTA are given for a comparison to a state-of-the-art PIV processing tool. Overall, the figure demonstrates an exceptional accuracy of WSSflow in providing physically correct flow quantities, especially with respect to the instantaneous wall-shear stress.  

Excellent agreement is observed between the streamwise velocity profiles obtained via WSSflow and the baseline profiles. This holds for the profiles averaged in time and streamwise direction (a,d) as well as an instantaneous profile at an arbitrary time step (b). Comparing the spatial resolution of the velocity profiles given by WSSflow with the output from the state-of-the-art PIV processing SOTA, the tremendous spatial resolution ability of WSSflow is striking. Furthermore, the capabilities of WSSflow allow to resolve the near-wall region (small $y^+$), which is of utmost importance for, e.g., understanding drag generation and turbulence production. On the contrary, SOTA is not able to represent the viscous sublayer ($y^+ \le 5$) nor can it accurately capture the velocity gradients in the buffer layer ($5 \lesssim y^+ \lesssim 30$) and the lower logarithmic layer ($y^+ \gtrsim 30$). This issue is widely known in the experimental fluid dynamics community~\citep{raffel_particle_2018,scharnowski2020} since it is rooted in the statistical approach of traditional PIV processing. By outputting a velocity estimate for each sub-window instead of each pixel like WSSflow, the original data resolution is massively coarsened. The inherent averaging across each sub-window further constrains the ability to resolve velocity gradients, which are particularly strong close to the wall. These limitations inhibit a derivation of the wall-shear stress from such data due to missing and/or unreliable velocity information in the viscous sublayer. Therefore, figure~\ref{fig:TCF}~(c,f) only depict the instantaneous wall-shear stress distributions of WSSflow. The synthetic configuration (c) follows the baseline remarkably well. Due to non-existing baseline information for real-world data, the experimental results (f) cannot be compared quantitatively. However, the magnitude and the frequency of the fluctuations is comparable to the synthetic data derived from high-fidelity numerical simulations, which indicates a physical significance in a qualitative sense.

Overall, the results related to the turbulent channel flow data build a valuable basis for future experimental analyses targeting, for instance, how different scales from the outer layer (large $y^+$) interact within a turbulent wall-bounded flow to modulate the wall-shear stress dynamics. Such analyses are currently limited to numerical investigations since the near-wall flow field cannot be sufficiently well resolved in space and time in experimental settings. On the contrary, high-fidelity numerical simulations are restricted to rather low Reynolds numbers due to the immense increase of the computational cost at high Reynolds numbers~\citep{vinuesa2022}. Hence, being able to resolve the instantaneous and spatially developing wall-shear stress experimentally is of utmost importance to understand the fundamental behavior of turbulent wall-bounded flows~\citep{mathis2013,maeteling2022} and how it can be actively manipulated to the benefit of society in the context of, e.g., friction drag reduction~\citep{marusic2021,maeteling2023}. For instance, in civil aviation, drag reduction of only a few percent can substantially reduce the required fossil fuels and the environmental burden~\citep{Ricco2021}. Consequently, our proposed framework could pave the way for a more sustainable and environmentally friendly transportation section.

\subsection{Wavy turbulent channel flow}\label{subsec:TWCF}
\begin{figure}[h]
   \centering
    \includegraphics[width=0.95\textwidth]{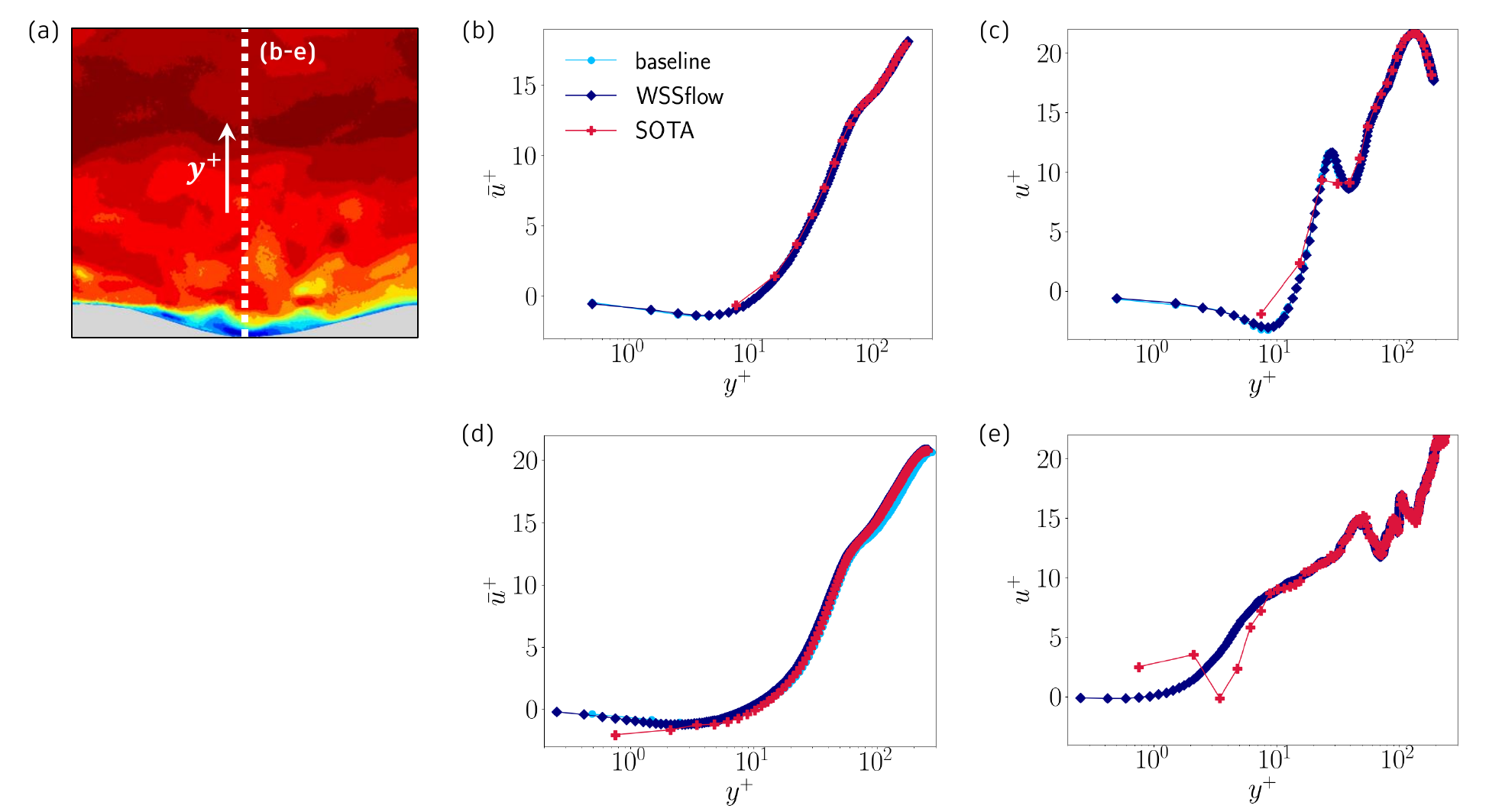}
    \caption{\textbf{Velocity distributions of the wavy turbulent channel flow data.} (a) shows an instantaneous flow field with a white dotted line that originates at the wave trough and indicates the position of the velocity profiles in (b-e). Velocity profiles in the upper row (b,c) are related to the synthetic particle images and the lower row (d,e) provides findings based on the experimental data. Note that these two datasets capture slightly different flows because the numerical data underlying the synthetic particle images has an adiabatic wall condition, while the experimental setup employs an isothermal boundary condition. (b,d) depict the time-averaged streamwise velocity profiles $\bar{u}^+$ as a function of the wall-normal distance $y^+$, where the $+$ sign denotes scaling in inner units. (c,e) show instantaneous velocity profiles $u^+$ of an arbitrary time step. Where baseline data are available for comparison (b-d), the velocity distributions of WSSflow follow the baseline exceptionally well. The velocity profiles of the traditional PIV processing tool SOTA have a much coarser spatial resolution, do not provide velocity information directly at the wall, and deviate severely from the true instantaneous distribution (c) in the buffer layer and below ($y^+ \lesssim 30$). Since the experimental data do not possess a known ground truth on an instantaneous level, the instantaneous velocity profiles (e) can only be validated qualitatively. However, the data provided by WSSflow is reasonably smooth, whereas the SOTA output contains non-physical jumps in the buffer layer and below.}
    \label{fig:TWCF_vel}
\end{figure}

\begin{figure}[h]
   \centering
    \includegraphics[width=0.95\textwidth]{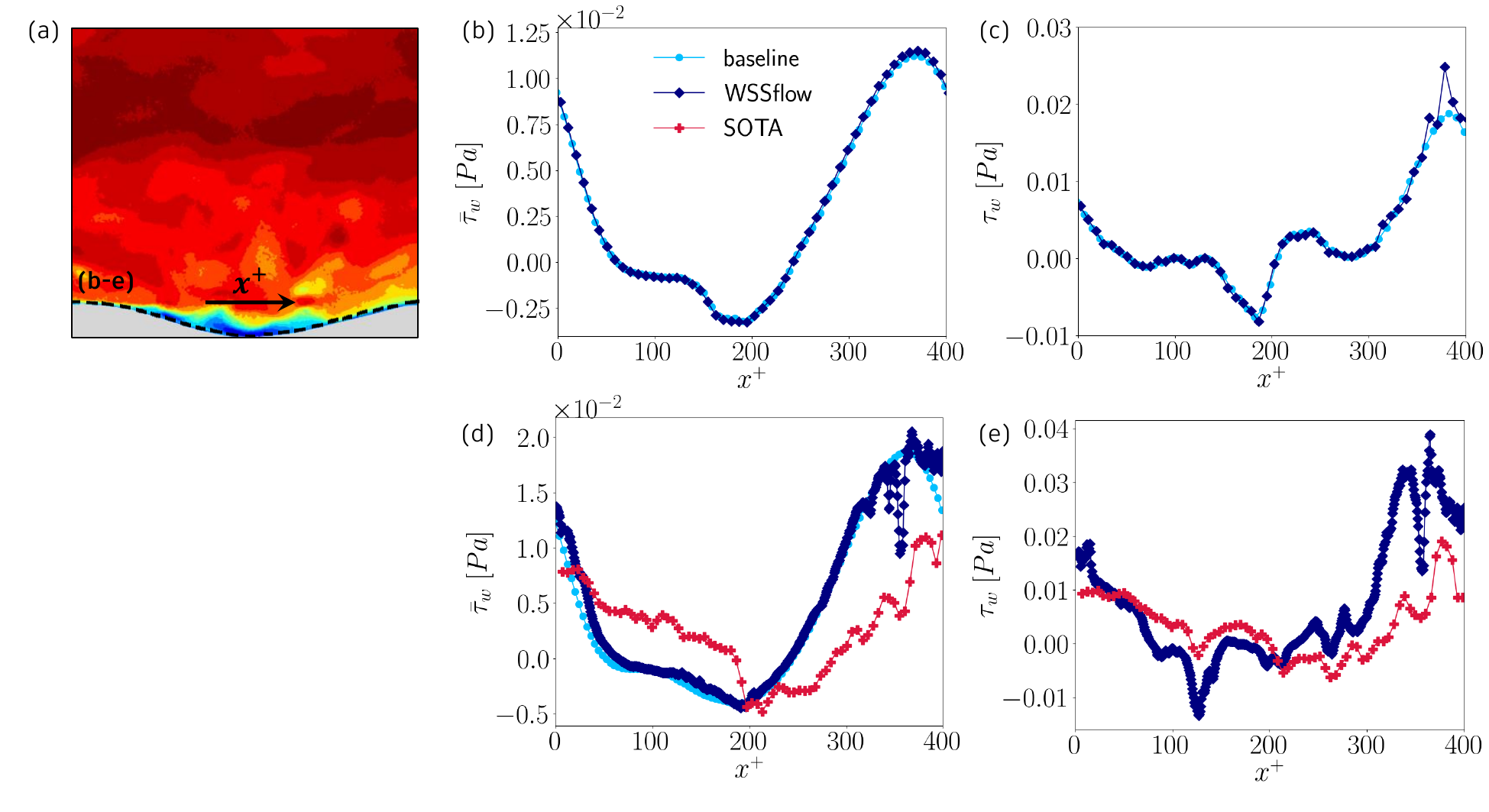}
    \caption{\textbf{Wall-shear stress distributions of the wavy turbulent channel flow data.} (a) shows an instantaneous flow field with a black dashed line that follows the wall contour and indicates the spatial extent of the wall-shear stress data given in (b-e). Wall-shear stress distributions in the upper row (b,c) are related to the synthetic particle images and the lower row (d,e) provides findings based on the experimental data. Note that these two datasets capture slightly different flows because the numerical data underlying the synthetic particle images has an adiabatic wall condition, while the experimental setup employs an isothermal boundary condition. (b,d) show the time-averaged wall-shear stress distributions $\bar{\tau}_w$ along the inner-scaled streamwise direction $x^+$. Only every 4th datapoint is shown for clarity. (c,e) present the instantaneous wall-shear stress distributions $\tau_w$ of an arbitrary time step. Both distributions obtained from the synthetic data (b,c) via WSSflow follow the baseline exceptionally well. Likewise, the time-averaged wall-shear stress estimated by WSSflow from the experimental data (d) accurately matches the baseline. Only one severe deviation occurs near the wave crest on the right side ($x^+ \approx 350$), which arises from laser light reflections at the wall that could not be avoided in the experimental setup. Since the experimental data do not possess a known ground truth on an instantaneous level, the instantaneous wall-shear stress distribution (e) can only be validated qualitatively. Nevertheless, the wall-shear stress fluctuation intensity is comparable to the synthetic data indicating a successful estimation. Note that although the spatial resolution of the experimental data is sufficiently high to provide wall-shear stress estimates using the traditional evaluation tool (SOTA), the obtained distributions are not physically meaningful for both, time-averaged (d) and instantaneous (e) quantities. This issue highlights the necessity of a novel framework like WSSflow that provides accurate physical flow quantities for detailed flow analyses.}
    \label{fig:TWCF_wss}
\end{figure}

Similar to the flat turbulent channel flow, synthetic and experimental particle images of the wavy turbulent channel flow are analyzed. Although the flow features exhibit more complex dynamics compared to the flat turbulent channel flow, the flow quantities extracted by WSSflow are similarly accurate. Instantaneous and time-averaged streamwise velocity (figure~\ref{fig:TWCF_vel}) and wall-shear stress (figure~\ref{fig:TWCF_wss}) distributions match the baseline data remarkably well.

As in the case of the turbulent channel flow, the flow quantities given by SOTA have a coarser spatial resolution than the neural counterpart. Moreover, when approaching the wall ($y^+ = 0$), the deviation of the velocity profiles from the baseline increases progressively as observed in figure~\ref{fig:TWCF_vel}. Even though SOTA does not capture the velocity values closest to the wall, the measurement setup is specifically designed to allow a detection of the viscous sublayer and the buffer layer even with classical PIV processing tools. Therefore, in contrast to the flat turbulent channel flow, a wall-shear stress estimation based on SOTA is possible. This enables a direct comparison between traditional and novel processing of the experimental dataset as

provided in figure~\ref{fig:TWCF_wss}~(d,e). The deviations between the streamwise evolution of the time-averaged and the instantaneous wall-shear stress determined by WSSflow and by SOTA are immense. For the time-averaged profile (d), WSSflow represents the baseline extraordinarily well except for the region near the wave crest ($x^+ \approx 350$). Laser light reflections at the wall cover the viscous sublayer in this area such that an accurate wall-shear stress determination fails. However, the satisfactory consistency with the baseline at the remaining locations demonstrates the success of the novel framework as long as measurement errors are minimized. On the contrary, the capability of SOTA to capture the true physics is severely limited. Neither the large-scale trend of the mean wall-shear stress (d) nor the instantaneous distribution (e) is represented precisely by this state-of-the-art PIV processing. Consequently, even if the viscous sublayer is captured by a traditional PIV evaluation routine as in the present setup, the severe velocity gradient smoothing inherent to this approach~\citep{raffel_particle_2018,scharnowski2020} inhibits an accurate estimation of the wall-shear stress.

The high-resolution experimental velocity and wall-shear stress distributions obtained via WSSflow are extremely important to understand how pressure gradients of varying strength impact the interaction of different flow scales. Even more relevant is the related analysis of how the pressure induced modification of the interaction phenomena affects the intensity and the dynamics of the wall-shear stress fluctuations, which can now conveniently be studied using the newly proposed framework WSSflow. Although such analyses are on a fundamental academic level, their findings impact a variety of real-world applications that rely on the efficiency of, e.g., transportation systems.

\subsection{Elastic blood vessel flow}\label{subsec:vessel}
\begin{figure}[h]
   \centering
    \includegraphics[width=0.99\textwidth]{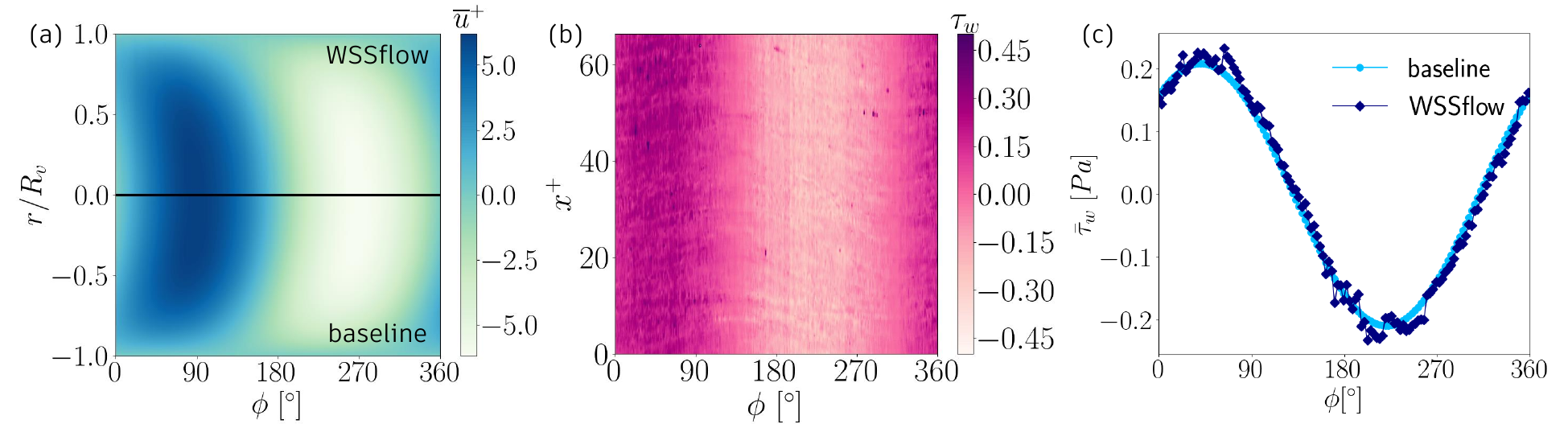}
    \caption{\textbf{Velocity and wall-shear stress distributions of the elastic blood vessel data.} (a)~shows color contours of the phase-averaged streamwise velocity as a function of the phase angle $\phi$ and the radial position $r/R_v$ with $R_v$ being the vessel radius. The upper part ($r/R_v > 0$) reflects the experimental data evaluated with WSSflow and the lower part ($r/R_v < 0$) is the baseline data. Since the phase-averaged flow is rotationally symmetric, positive and negative radial directions are equivalent. (b)~provides color contours of the instantaneous wall-shear stress distribution of the experimental data obtained via WSSflow as a function of $\phi$ and the streamwise position $x^+$. (c) presents the phase-averaged experimental and baseline wall-shear stress distributions as a function of the phase angle $\phi$. Only every 4th datapoint of the WSSflow prediction is shown for clarity. The experimentally obtained phase-averaged velocity distribution matches the baseline very well (a). Similar consistency occurs for the phase-averaged wall-shear stress distribution (c) with minor deviations at the peak values, which demonstrates the fluid-structure interaction. The observed similarities indicate that also the instantaneous wall-shear stress distribution (b), which is an interim result for obtaining the phase-averaged data, must be of physical significance.}
    \label{fig:vessel}
\end{figure}

We are considering the oscillatory flow through a transparent elastic blood vessel model at flow conditions similar to the human abdominal aorta. The streamwise velocity field and the wall-shear stress distributions obtained by evaluating the experimental particle images with WSSflow are shown in figure~\ref{fig:vessel}. The phase-averaged velocity field is shown in the upper part of figure~\ref{fig:vessel}~(a) in dependence of the radial position and the phase angle, whereas the baseline data of a rigid vessel is provided in the lower part. For this representation, instantaneous velocity fields are averaged in the streamwise direction and over six oscillation cycles. Both distributions appear very similar because of the small elastic vessel dilatation of $0.4~\%$ with respect to the rigid vessel diameter. Relative to the vessel wall thickness, the wall movement amounts to $8.3~\%$.

An instantaneous wall-shear stress distribution along the streamwise direction and in dependence of the phase angle (b) shows that the wall-shear stress possesses a similar phase dependence than the velocity field but with a phase shift relative to the velocity. Minor variations in the streamwise direction are observed, which highlight the importance of instantaneous spatially developing wall-shear stress distributions for detailed flow analysis, e.g., with respect to the arterial disease development. Applying a spatial and phase average to all wall-shear stress fields (c) allows a comparison to the baseline wall-shear stress of a rigid vessel flow, which shows that minor deviations primarily occur at absolute peak values. These emanate from the fluid-structure interaction between the blood flow and the elastic vessel. The non-smoothness of the experimental data originates from the fact that only six oscillation cycles are phase-averaged.\\

Overall, these findings show that a wall-shear stress estimation from particle images of a real-world relevant experimental setup is remarkably accurate when using WSSflow. This bears enormous potential for experimental studies targeting the impact of the wall-shear stress dynamics on health related conditions such as atherosclerosis, where spatially resolved wall-shear stress data on an instantaneous level is key to understand the time and space dependent phenomena and develop sophisticated prevention and treatment strategies. 

\section{Discussion and conclusion}\label{sec:discussion}
In this paper, we demonstrate that the combination of experimental PIV measurements with cutting-edge deep learning methods allows for practical and accurate approaches to derive fluid flow quantities that are typically difficult - in most cases even impossible - to measure. Precisely, we propose a novel end-to-end workflow based on \textit{WSSflow} which determines highly resolved velocity and wall-shear stress information from widely used planar PIV measurements.  
In contrast to traditional PIV evaluation, a velocity field is provided at the original image resolution. Leveraging physical knowledge, our framework WSSflow then calculates the wall-shear stress from the near-wall velocity distribution in an instantaneous and spatially resolved fashion.  
That is, no further assumptions or modeling hypotheses, which typically underlie complementary approaches, are deployed. 

The success of the proposed framework is demonstrated using synthetic and experimental datasets of three test cases. In particular, the synthetic particle images of a flat and a wavy turbulent channel flow are used to verify that the estimated temporally and spatially resolved wall-shear stress distributions accurately follow the underlying baseline data. The application to experimental measurement data additionally including a realistic oscillating blood flow through an elastic vessel clearly demonstrate a physical significance of the estimated flow quantities even for particle images containing typical measurement uncertainties. Comparisons to a state-of-the-art PIV evaluation show the distinct superiority of the novel framework WSSflow with respect to spatial resolution, the ability to resolve local velocity gradients, the ability to capture near-wall physics, and the overall accuracy of the estimated flow quantities. Thus, it can be concluded that WSSflow constitutes a potential game changer for experimental fluid dynamics. It allows to derive an extremely important flow quantity, i.e., the wall-shear stress, with an outstanding spatial and temporal resolution as it is required for detailed flow analysis in both, academia and real-world related disciplines. For instance, our method can be used to experimentally study novel friction drag reduction techniques for the transportation sector or the mechanisms by which cardiovascular diseases develop and how these phenomena can be attenuated to prevent vascular dysfunction. 

Since the flow data obtained via WSSflow possess a very high spatial resolution, they can match or even surpass the spatial resolution of high-fidelity large-scale simulations. Consequently, a key direction for future work will be to validate and benchmark these exceptionally costly simulations using the proposed workflow. In follow-up work, it would be interesting to exploit the full potential of our approach via a direct coupling with numerical simulations using data assimilation techniques. That is, future models leveraging our workflow pipeline might be used to derive reduced-order or parameterized model formulations that can be directly queried from numerical solvers. This would amount in an immense reduction of the computational cost with the tremendous benefit of experimentally validated and/or derived models. Another promising future direction is the introduction of physically motivated inductive biases. Extending the physical knowledge provided to the learner has the potential to enhance the measurement data in different aspects as demonstrated on the basis of Physics-Informed Neural Networks (PINNs)~\cite{raissi2019} which are based on a Navier-Stokes equations regularized loss. Especially in regions of marginal texture, immense occlusion, or overexposure due to reflections, physically motivated regularizers and latent dynamical models \cite{lagemann2023learning} might be able to provide otherwise unavailable flow field information and could also infer additional flow quantities like temperature and pressure. Moreover, a direct coupling with causal risk frameworks \cite{lagemann2022deep} is very interesting to additionally extract causal relations in the underlying flow. 

\section{Methods}\label{sec:methods}
In this section, we provide information about the proposed approach, the evaluation scheme, and a detailed presentation of the applied architecture and associated implementation.\\

\textbf{Notation.} 
Let $\boldsymbol{X}$ be the input of input size $(T \times M \times N \times C)$. Entries are denoted $ \boldsymbol{X}_{t,m,n,c}$ with $t$ being the snapshot index in the temporal domain, $m,n$ the spatial indices in the horizontal and vertical image direction, and $c$ the differentiation between the first and second particle image. Occasionally, we overload this notation by a simplified indexing when the entire set of first or second images is used, hence $\boldsymbol{X}_1$ denotes all $T$ images in $\boldsymbol{X}$ or formally $\boldsymbol{X}_1 = \boldsymbol{X}_{:,:,:,1}$. Since training and inference datasets correspond to different image feature distributions (which necessarily do not match), training data is indicated via $\boldsymbol{X}^\ast$ while inference data used for post-processing is denoted by $\boldsymbol{X}$ (further details appear below). 
During inference, our neural optical flow framework WSSflow is employed to output optical flows for all $T$ image pairs in $\boldsymbol{X}$. Optical flow is denoted by $\boldsymbol{V} \in \mathbb{R}^{m \times n \times 2} = (V_1, V_2)$ with $V_1, V_2$ being the horizontal and vertical optical flow component, respectively. Thus, the general optical flow problem is formalized as $\boldsymbol{V} = F_\theta (\boldsymbol{X})$, with $F$ being any arbitrary non-linear but learnable mapping and $\theta$ covering the entire set of trainable parameters in WSSflow. Specific network weights are given by $\boldsymbol{w}$ whose dimensions become clear from the context. Non-trainable network entities such as the correlation volume $\boldsymbol{\mathcal{C}}$ are introduced when necessary. $\mathcal{L}$ refers to the loss minimizing the loss objective during training. Relevant fluid mechanical quantities are introduced occasionally when necessary and become clear from the context. \\

\textbf{Problem statement.}
We focus on the setting in which available inputs are
\begin{itemize}
\item[] (I1) particle image pairs as an $(T \times M \times N \times C)$ matrix whose first dimension corresponds to measurement snapshots, whose second and third dimension are spatial dimensions in streamwise and wall-normal direction, and whose last dimension represents two individual particle images separated by a small interframing time $\Delta t$. 
\item[] (I2) length and time scale knowledge of the measurement setup providing information to calibrate pixel-based image features yielding real-world velocity data.
\end{itemize}
For (I1), we assume that the particle images cover a broad range of flow scales simultaneously comprising the large scales in the outer flow and the near-wall small-scale flow structures. Precisely, we require the measurement setup to resolve the viscous sublayer by at least one pixel in camera coordinates. Complementary, a sufficient number of particles is required to predict the displacement distribution reliably. In fact, this holds similarly for wall-distant regions such as the buffer layer and beyond. For (I2), we assume precise calibration information to convert pixel-based image features to real-world flow information. Hence, the interframing time $\Delta t$ and a pixelwise calibration model is expected to be known. Note that a pixelwise wall location needs to be accessible, too. Moreover, fluid properties such as temperature, density, and viscosity are required. \\
Suppose all input requirements are fulfilled, the underlying learning task of our approach can be formulated as follows: Given the inputs (I1) and (I2), our goal is to predict a pixelwise optical flow field for each individual image pair in $\boldsymbol{X}$ and derive highly relevant but typically difficult to measure near-wall flow quantities such as the wall-shear stress $\tau_w$ over the entire temporal and spatial domain captured in $\boldsymbol{X}$. \\

\textbf{Learning scheme.}
With the notation given above, our goal is to learn an optical flow estimator predicting high-resolution displacement distributions of recorded particle image pairs. To this end, we train a parameterized network $F_\theta$, i.e. a non-linear function $F$ with a set of unknown trainable parameters $\theta$. This is possible since training is performed in a supervised fashion assuming a carefully designed synthetic dataset with known ground-truth labels. Note that training is performed prior and independently to inference on $\boldsymbol{X}$. Hence, we require our network $F_\theta$ to generalize well to \textit{unknown} out-of-distribution real-world image features in an out-of-the-box fashion making training particularly important for the overall velocity and wall-shear stress derivation. The architecture we use for $F_\theta$ is detailed below, but for now assume that this has been specified. Then, given the training data $\boldsymbol{X}^\ast$ (which is independent in feature and flow distribution from our inference data $\boldsymbol{X}$) and the training labels $\boldsymbol{Y}^\ast$ for all image pairs $\boldsymbol{X}^\ast$, we train the set of parameters $\hat{\theta}(\boldsymbol{X}^\ast)$ under a loss that is supervised by the labels $\boldsymbol{Y}^\ast$. In our setting, we consider a $L_1$-loss between ground truth and optical flow estimate for all image pairs $\boldsymbol{X}^\ast$, where the ground truth provides the real optical flow field (see details below). Assuming a suitable training dataset $\boldsymbol{X}^\ast$ with corresponding labels $\boldsymbol{Y}^\ast$, minimizing this $L_1$-loss during training prevents exploding weights by default and enables sufficient generalization capabilities as demonstrated empirically. \\

\textbf{Architectural details of WSSflow.}
The proposed framework WSSflow is an extended version of RAFT-PIV \cite{lagemann2021}, which is a neural optical flow estimator based on the successful RAFT architecture \cite{teed2020} and specifically designed for the use case of PIV images. It is unique in the sense that it operates entirely on a specific input resolution and updates iteratively its flow predictions. \\
WSSflow consists of the following stages: a feature and a context extracting block, the computation of a pixelwise correlation volume using an all-pairs correlation, a GMA, iterative updates based on a ConvGRU, and a physics informed transformation as shown in Figure~\ref{fig:overview}. In a first step, the shared feature encoder derives latent embeddings $\boldsymbol{E_1}$, $\boldsymbol{E_2}$ of size ($M \times N \times D$) for each input image individually using three convolutional neural network modules. These encoding steps map the input particle images into a dense feature representation with $D$ dimensions. On an abstract level, these feature map represent an extract of the most relevant image patterns like edges or simple textures while deeper convolutional filters capture more complex relations or even entire objects. To this end, the neural optical flow estimator is trained to predict the local displacement between two images based on the detection of various patterns.\\
The second stage of WSSflow is designed to compute the similarity of both image features $\boldsymbol{E}_1$ and $\boldsymbol{E}_2$ using a full correlation volume between all spatial indices of both feature maps. The similarity between two pixel embeddings ($\textbf{E}_{i,j}, \textbf{E}_{k,l}$) is measured by the dot product between the two individual feature vectors yielding the so-called correlation volume
\begin{equation}
    \boldsymbol{\mathcal{C}}_{ijkl} = \sum_{d=1}^D (\boldsymbol{E}_1)_{ijd} \cdot (\boldsymbol{E}_2)_{kld}
\end{equation}
with $i,j$ denoting pixel coordinates in the image embedding $\boldsymbol{E}_1$ and $k,l$ in $\boldsymbol{E}_2$. Here, all-pairs correlation means that every pixel is correlated with every other pixel. Hence, $\boldsymbol{\mathcal{C}}(i,j,:,:)$ represents a correlation map of pixel $(i,j)$ in $\boldsymbol{E}_1$ with all pixels of the second image $\boldsymbol{E}_2$. Based on this key idea, a 4-layer correlation pyramid is formed (see Figure \ref{fig:overview}). Precisely, the last two dimensions of $\boldsymbol{\mathcal{C}}$  are pooled sequentially from level to level using kernel sizes and strides of ${1, 2, 4, 8}$. Thus, WSSflow maintains high resolution information of the first image while effectively addressing large object displacements applying pooling operations along the features of the second image.
Note that this correlation volume is only computed once in advance and only small neighborhoods of a specific target location are extracted during the update process. For clarification, let us assume that the current optical flow of the $i$-th update step is available. Now, each pixel in $\boldsymbol{E}_1$ is mapped to its estimated location in $\boldsymbol{E}_2$. To obtain the valid index of the correlation volume, the index operation considers a specific neighbourhood around the estimated location in $\boldsymbol{E}_2$ and is applied to all levels of the correlation pyramid using a constant radius of $4 \,\mathrm{px}$ on the neighbourhood grid. Subsequently, the resulting values of each level are concatenated to a single feature map.
The idea behind this complex index operation can be formulated as follows: for a specific target location $(i,j)$ in encoding $\boldsymbol{E}_1$, the corresponding value in the correlation volume expresses the similarity of a pixel embedding at position $(k,l)$ in encoding $\boldsymbol{E}_2$ and its level-specific neighbourhood. Hence, the higher the pyramid level of the extracted patch, the more spatial information of the original image features is covered, but the coarser the resolution of the correlation information becomes. This multi-scale approach allows WSSflow to resolve large displacements of small objects quite accurately.

In this work, we augment WSSflow by a global motion aggregation (GMA) module similar to the one proposed in \cite{jiang2021learning}. It operates on the combined input of the correlation pyramid output and the context encoded first particle image. The idea behind this component can be formulated as follows:
The knowledge of linked pixels with high similarity and the corresponding feature embeddings in non-occluded regions can be transferred to occluded image parts.
To do so, the GMA considers long-range feature connections extracted by a self-attention mechanism and estimates motion features which are propagated to occluded regions. Its main design is strongly related to the recently famous self-attention mechanism used in transformer networks \cite{dosovitskiy2020image}. From a high-level perspective, the GMA utilizes attention in a specific way to introduce the notion of memory over time. Precisely, the attention weights indicate the relevance in time and pixel space.
Unlike convolutions whose receptive field is limited to a fraction of the input sequence, the self-aware attention formulation principally grants access to all parts of the entire input sequence.
As a result, all pixel embeddings can be considered simultaneously and the network learns to pay attention to the most relevant features depending on the task at hand.\\
From a technical point of view, self-attention can be computed as follows: Let $\boldsymbol{E}_q \in \mathbb{R}^{(M\cdot N) \times D_q}$ be the context feature map, $\boldsymbol{E}_{q,i} \in \mathbb{R}^{D_q }$ is its $i$-th feature vector, and $\textbf{E}_r \in \mathbb{R}^{(M \cdot N) \times D_r}$ denotes the motion features captured in the correlation volume. Then, the aggregated motion features can be expressed by the self-aware linear recombination of the query, key and value vector as follows \cite{jiang2021learning}:
\begin{equation}
    \hat{\boldsymbol{E}}_r = \boldsymbol{E}_r + \alpha \sum_{j=1}^{(M \cdot N)} \mathcal{F} \left (  \Theta(\boldsymbol{E}_{q,i}), \Phi(\boldsymbol{E}_{q,j})\right ) \zeta(\boldsymbol{E}_{r,j}),
\end{equation}
where $\alpha$ is a learned parameter and $\Theta$, $\Phi$, and $\zeta$ denote the projection of the query, key and value vectors which are given for the $i$-th and $j$-th feature vector by 
\begin{eqnarray}
\Theta(\boldsymbol{E}_{q,i}) &=& \boldsymbol{w}_{Q}\boldsymbol{E}_{q,i}\\
\Phi(\boldsymbol{E}_{q,j}) &=& \boldsymbol{w}_{K}\boldsymbol{E}_{q,j}\\
\zeta(\boldsymbol{E}_{r,j}) &=& \boldsymbol{w}_{V}\boldsymbol{E}_{r,j} .
\end{eqnarray}

The function $\mathcal{F}$ represents the similarity attention function \cite{vaswani2017attention} and reads
\begin{equation}
    \mathcal{F}(\textbf{a}_i, \textbf{b}_j) = \frac{\exp \left ( \textbf{a}_i^T \textbf{b}_j / \sqrt{D_q}\right )}{\sum_{j=1}^{(M \cdot N)} \exp \left ( \textbf{a}_i^T \textbf{b}_j / \sqrt{D_q}\right )} .
\end{equation}

The final output is then a concatenation of the previous hidden state, the general motion features, and the aggregated motion features. It is subsequently decoded in the ConvGRU~\citep{siam2017convolutional} to obtain a new optical flow update, which basically allows a combination of motion vectors and self-aware context features.\\
A ConvGRU is a type of recurrent neural network that stores hidden state information of previous steps to modulate a limited content memory. Thus, observations of previous steps are taken into account when estimating future predictions. The main difference to a standard recurrent neural network is the so-called update and reset gate. The applied ConvGRU takes flow, correlation, and the hidden state of the context network as input and yields a new hidden state $\textbf{h}_t$. Subsequently, $\textbf{h}_t$ is passed through two convolutional layers and finally outputs the flow update $\Delta \boldsymbol{V}$. Thus, the final flow prediction is a combination of the sequence of residual flow updates. The benefit of a ConvGRU to perform the iterative refinement lies in the reduction of the search space due to its recurrent nature. This ConvGRU allows the network to balance the prediction of earlier optical flows and its current hidden state to compute a new optical flow update.\\
The final output of the ConvGRU, i.e., the optical flow, is a high-resolution velocity field at the original image resolution. Precisely, WSSflow estimates one velocity vector for each pixel of the particle image pair. Moreover, the velocity information is combined with physical knowledge to calculate the space-dependent wall-shear stress. We use the fact that the flow in the viscous sublayer, which is a very thin wall-parallel layer adjacent to the boundary with $y^+ \le 5$, is dominated by viscous forces. Therefore, the inner-scaled streamwise velocity $u^+ = u/u_{\tau}$ equals the inner-scaled wall-normal position $y^+ = \rho u_\tau y/\eta$, where $u_\tau$ is the friction velocity, $\rho$ the fluid density, and $\eta$ the dynamic viscosity of the fluid. Since the wall-shear stress $\tau_w$ is directly related to the friction velocity $\tau_w = \rho u_\tau^2$, it follows from $u^+ = y^+$ that $\tau_w = \eta u/y$. In other words, we deploy the linearity of the velocity gradient in the viscous sublayer. This allows us to estimate the wall-shear stress dynamics as a function of time and space, i.e., one wall-shear stress value is obtained for each streamwise location and each time instant. The only prerequisite for this approach is that the measurement setup captures the viscous sublayer by at least one pixel in camera coordinates. If more pixels cover the viscous sublayer, all velocity information in the viscous sublayer is used to calculate a wall-shear stress value as an average across the wall-normal direction. Our studies showed that this typically results in more accurate wall-shear stress estimates. Thus, we recommend a minimum number of three pixels in the viscous sublayer which has proven to outweigh measurement uncertainties arising directly at the wall, e.g., laser light reflections and a misalignment of the wall boundary on a sub-pixel level. \\

\textbf{Training details of WSSflow.}
 All network architectures are implemented in the open source framework PyTorch \cite{adam_paszke_automatic_2017}. All modules are initialized from scratch using random weights. During training, an AdamW-Optimizer \cite{loshchilov2018fixing} is applied starting at an initial learning rate $\varepsilon_0=0.0001$. Gradients are clipped to the range [-1,1]. Furthermore, the learning rate is reduced by a factor of five once the evaluation metrics stopped improving for 15 consecutive epochs. The minimum learning rate is set to  $\varepsilon_{min}=10^{-8}$. This learning rate scheduler shows the best results in the current study.\\ 
 In total, $200$ epochs are used for training. All computations are run on a single GPU node equipped with four NVIDIA A100 resulting in an overall training time of approximately $22 \,\mathrm{h}$ for a global batch size of 8. With WSSflow operating exclusively on the initial input resolution, a training image size of $128 \times 128 \,\mathrm{px^2} $ is chosen to bypass extreme memory consumption when computing the all-pairs correlation volume. To overcome this issue during inference, evaluation is performed on subpatches of each image pair which are stitched together in a final step. \\

\textbf{Principles of PIV and traditional PIV evaluation routine SOTA.} 
For visualization purposes, PIV requires the addition of tracer particles to the flow. The particles' properties are matched to the flow characteristics such that the particle motion closely follows the flow dynamics. Using a pulsed laser light sheet that is synchronized with a high-quality camera, particle images are acquired. Based on the displacement of the tracer particles between two consecutive images and their interframing time, the velocity distribution can be derived. In traditional PIV processing, this is performed on a statistical basis~\citep{raffel_particle_2018}. Therefore, the original particle image is divided into finite-size interrogation windows which are cross-correlated. The position of the correlation peak indicates the averaged displacement within the respective image section. This approach substantially lowers the spatial resolution of the resulting velocity field compared to the original image size. For instance, for a particle image pair of size $1024 \times 1024$~px$^2$, which might typically be evaluated with an interrogation window size of $32 \times 32$~px$^2$ and a maximum window overlap of $75$~\%, only $124 \times 124$ velocity vectors are obtained. Furthermore, the spatial averaging involved in such statistical approaches smooths velocity gradients within each interrogation window.\\
The proposed framework WSSflow is benchmarked against a high-performance in-house code \cite{marquardt2019, marquardt2020experimental} which represents the current gold-standard of PIV evaluation algorithms and is referred to as SOTA. First, a 4-level multi-grid evaluation with an initial window size of $256 \times 256\,\mathrm{px}^2$ and a final window size of $32 \times 32\,\mathrm{px}^2$ is performed. This enables the detection of displacements larger than half of the interrogation window size with successive refinement. Subsequently, a five-step predictor-corrector is used in an iterative fashion. Here, a subpixel accuracy is achieved by applying the approach of \cite{astarita2005analysis}. A second-order accurate displacement estimate is derived by shifting and deforming the interrogation windows by half of the initially computed displacement featuring a B-Spline-3 interpolation to obtain a per-pixel accurate displacement field during window deformation. Image interpolation is based on a 4th-order Lanczos interpolation \cite{gallivan1996rational}. Following \cite{schrijer_effect_2008}, an integral displacement predictor is used to ensure convergence during iterations yielding a weighted average of the per-pixel displacement field over the entire subwindow. The complementary corrector is designed according to \cite{raffel_particle_2018}. That is, the cross-correlation between corresponding interrogation windows of size $32 \times 32\,\mathrm{px}^2$ (with $75\%$ overlap) is computed using a 3-point Gaussian peak estimator. The windows are weighted by a 2D Gaussian with a standard deviation of $\sigma=0.4$. Outlier detection is performed using a normalized median test and the respective outliers are replaced by an interpolation of the neighboring $3 \times 3$ vectors. 

\bibliographystyle{plain}

\begin{thebibliography}{10}

\bibitem{abbassi2017}
MR~Abbassi, Woutijn~J Baars, Nicholas Hutchins, and Ivan Marusic.
\newblock Skin-friction drag reduction in a high-{R}eynolds-number turbulent boundary layer via real-time control of large-scale structures.
\newblock {\em International Journal of Heat and Fluid Flow}, 67:30--41, 2017.

\bibitem{adamo2009biomechanical}
Luigi Adamo, Olaia Naveiras, Pamela~L Wenzel, Shannon McKinney-Freeman, Peter~J Mack, Jorge Gracia-Sancho, Astrid Suchy-Dicey, Momoko Yoshimoto, M~William Lensch, Mervin~C Yoder, et~al.
\newblock Biomechanical forces promote embryonic haematopoiesis.
\newblock {\em Nature}, 459(7250):1131--1135, 2009.

\bibitem{albers2021}
Marian Albers and Wolfgang Schr{\"o}der.
\newblock Lower drag and higher lift for turbulent airfoil flow by moving surfaces.
\newblock {\em International Journal of Heat and Fluid Flow}, 88:108770, 2021.

\bibitem{alfredsson1988}
P~Henrik Alfredsson, Arne~V Johansson, Joseph~H Haritonidis, and Helmut Eckelmann.
\newblock The fluctuating wall-shear stress and the velocity field in the viscous sublayer.
\newblock {\em The Physics of Fluids}, 31(5):1026--1033, 1988.

\bibitem{arun2023}
Rahul Arun, H~Jane Bae, and Beverley~J McKeon.
\newblock Towards real-time reconstruction of velocity fluctuations in turbulent channel flow.
\newblock {\em Physical Review Fluids}, 8(6):064612, 2023.

\bibitem{arzani2021}
Amirhossein Arzani, Jian-Xun Wang, and Roshan~M D'Souza.
\newblock Uncovering near-wall blood flow from sparse data with physics-informed neural networks.
\newblock {\em Physics of Fluids}, 33(7), 2021.

\bibitem{astarita2005analysis}
Tommaso Astarita and Gennaro Cardone.
\newblock Analysis of interpolation schemes for image deformation methods in {PIV}.
\newblock {\em Experiments in Fluids}, 38(2):233--243, 2005.

\bibitem{bae2022}
H~Jane Bae and Petros Koumoutsakos.
\newblock Scientific multi-agent reinforcement learning for wall-models of turbulent flows.
\newblock {\em Nature Communications}, 13(1):1443, 2022.

\bibitem{balasubramanian2023}
Arivazhagan~G. Balasubramanian, Luca Guastoni, Philipp Schlatter, Hossein Azizpour, and Ricardo Vinuesa.
\newblock Predicting the wall-shear stress and wall pressure through convolutional neural networks.
\newblock {\em International Journal of Heat and Fluid Flow}, 103:109200, 2023.

\bibitem{bellien2021}
J{\'e}r{\'e}my Bellien, Michele Iacob, Vincent Richard, Julien Wils, Veronique Le~Cam-Duchez, and Robinson Joannid{\`e}s.
\newblock Evidence for wall shear stress-dependent t-{PA} release in human conduit arteries: role of endothelial factors and impact of high blood pressure.
\newblock {\em Hypertension Research}, 44(3):310--317, 2021.

\bibitem{boris1992}
Jay~P Boris, Fernando~F Grinstein, Elaine~S Oran, and Ronald~L Kolbe.
\newblock New insights into large eddy simulation.
\newblock {\em Fluid Dynamics Research}, 10(4-6):199, 1992.

\bibitem{brown2016}
Adam~J Brown, Zhongzhao Teng, Paul~C Evans, Jonathan~H Gillard, Habib Samady, and Martin~R Bennett.
\newblock Role of biomechanical forces in the natural history of coronary atherosclerosis.
\newblock {\em Nature Reviews Cardiology}, 13(4):210--220, 2016.

\bibitem{brunton2020}
Steven~L Brunton, Bernd~R Noack, and Petros Koumoutsakos.
\newblock Machine learning for fluid mechanics.
\newblock {\em Annual Review of Fluid Mechanics}, 52:477--508, 2020.

\bibitem{cai2019}
Shengze Cai, Shichao Zhou, Chao Xu, and Qi~Gao.
\newblock Dense motion estimation of particle images via a convolutional neural network.
\newblock {\em Experiments in Fluids}, 60(4):73, 2019.

\bibitem{chamorro2013}
Leonardo~P Chamorro, REA Arndt, and Fotis Sotiropoulos.
\newblock Drag reduction of large wind turbine blades through riblets: Evaluation of riblet geometry and application strategies.
\newblock {\em Renewable Energy}, 50:1095--1105, 2013.

\bibitem{chung2021natural}
Min~Gon Chung, Kenneth~A Frank, Yadu Pokhrel, Thomas Dietz, and Jianguo Liu.
\newblock Natural infrastructure in sustaining global urban freshwater ecosystem services.
\newblock {\em Nature Sustainability}, 4(12):1068--1075, 2021.

\bibitem{clauser1956}
Francis~H Clauser.
\newblock The turbulent boundary layer.
\newblock {\em Advances in Applied Mechanics}, 4:1--51, 1956.

\bibitem{colella2003}
Kurt~J Colella and William~L Keith.
\newblock Measurements and scaling of wall shear stress fluctuations.
\newblock {\em Experiments in Fluids}, 34:253--260, 2003.

\bibitem{cooper2004}
Kevin~R Cooper.
\newblock Commercial vehicle aerodynamic drag reduction: historical perspective as a guide.
\newblock In {\em The Aerodynamics of Heavy Vehicles: Trucks, Buses, and Trains}, pages 9--28. Springer, 2004.

\bibitem{dalili2009}
Neda Dalili, Afsaneh Edrisy, and Rupp Carriveau.
\newblock A review of surface engineering issues critical to wind turbine performance.
\newblock {\em Renewable and Sustainable Energy Reviews}, 13(2):428--438, 2009.

\bibitem{dosovitskiy2020image}
Alexey Dosovitskiy, Lucas Beyer, Alexander Kolesnikov, Dirk Weissenborn, Xiaohua Zhai, Thomas Unterthiner, Mostafa Dehghani, Matthias Minderer, Georg Heigold, Sylvain Gelly, et~al.
\newblock An image is worth 16x16 words: Transformers for image recognition at scale.
\newblock In {\em The Ninth International Conference on Learning Representations}, 2021.

\bibitem{duan2020}
Huan-Feng Duan, Bin Pan, Manli Wang, Lu~Chen, Feifei Zheng, and Ying Zhang.
\newblock State-of-the-art review on the transient flow modeling and utilization for urban water supply system ({UWSS}) management.
\newblock {\em Journal of Water Supply: Research and Technology-Aqua}, 69(8):858--893, 2020.

\bibitem{fernholz1996}
Hans-Hermann Fernholz and PJ~Finley.
\newblock The incompressible zero-pressure-gradient turbulent boundary layer: an assessment of the data.
\newblock {\em Progress in Aerospace Sciences}, 32(4):245--311, 1996.

\bibitem{franke1984induction}
R-P Franke, M~Gr{\"a}fe, H~Schnittler, D~Seiffge, Ch~Mittermayer, and D~Drenckhahn.
\newblock Induction of human vascular endothelial stress fibres by fluid shear stress.
\newblock {\em Nature}, 307(5952):648--649, 1984.

\bibitem{fureby1999}
Christer Fureby and Fernando~F Grinstein.
\newblock Monotonically integrated large eddy simulation of free shear flows.
\newblock {\em AIAA journal}, 37(5):544--556, 1999.

\bibitem{gallivan1996rational}
Kyle Gallivan, G~Grimme, and Paul Van~Dooren.
\newblock A rational {Lanczos} algorithm for model reduction.
\newblock {\em Numerical Algorithms}, 12(1):33--63, 1996.

\bibitem{graham2016}
Jason Graham, K~Kanov, XIA Yang, M~Lee, N~Malaya, CC~Lalescu, R~Burns, G~Eyink, A~Szalay, RD~Moser, et~al.
\newblock A web services accessible database of turbulent channel flow and its use for testing a new integral wall model for {LES}.
\newblock {\em Journal of Turbulence}, 17(2):181--215, 2016.

\bibitem{grant2012taking}
Stanley~B Grant, Jean-Daniel Saphores, David~L Feldman, Andrew~J Hamilton, Tim~D Fletcher, Perran~LM Cook, Michael Stewardson, Brett~F Sanders, Lisa~A Levin, Richard~F Ambrose, et~al.
\newblock Taking the “waste” out of “wastewater” for human water security and ecosystem sustainability.
\newblock {\em Science}, 337(6095):681--686, 2012.

\bibitem{gubian2019}
Pierre-Alain Gubian, Jordan Stoker, James Medvescek, Laurent Mydlarski, and B~Rabi Baliga.
\newblock Evolution of wall shear stress with reynolds number in fully developed turbulent channel flow experiments.
\newblock {\em Physical Review Fluids}, 4(7):074606, 2019.

\bibitem{hartman2021}
Eline~MJ Hartman, Giuseppe De~Nisco, Frank~JH Gijsen, Suze-Anne Korteland, Anton~FW van~der Steen, Joost Daemen, and Jolanda~J Wentzel.
\newblock The definition of low wall shear stress and its effect on plaque progression estimation in human coronary arteries.
\newblock {\em Scientific Reports}, 11(1):22086, 2021.

\bibitem{ilg2017}
Eddy Ilg, Nikolaus Mayer, Tonmoy Saikia, Margret Keuper, Alexey Dosovitskiy, and Thomas Brox.
\newblock Flownet 2.0: Evolution of optical flow estimation with deep networks.
\newblock In {\em Proceedings of the IEEE conference on computer vision and pattern recognition}, pages 2462--2470, 2017.

\bibitem{jiang2021learning}
Shihao Jiang, Dylan Campbell, Yao Lu, Hongdong Li, and Richard Hartley.
\newblock Learning to estimate hidden motions with global motion aggregation.
\newblock In {\em Proceedings of the IEEE/CVF International Conference on Computer Vision}, pages 9772--9781, 2021.

\bibitem{kahler2012}
Christian~J K{\"a}hler, Sven Scharnowski, and Christian Cierpka.
\newblock On the uncertainty of digital {PIV} and {PTV} near walls.
\newblock {\em Experiments in Fluids}, 52:1641--1656, 2012.

\bibitem{kahler2006}
Christian~J K{\"a}hler, U~Scholz, and J~Ortmanns.
\newblock Wall-shear-stress and near-wall turbulence measurements up to single pixel resolution by means of long-distance micro-{PIV}.
\newblock {\em Experiments in Fluids}, 41:327--341, 2006.

\bibitem{kim1987turbulence}
John Kim, Parviz Moin, and Robert Moser.
\newblock Turbulence statistics in fully developed channel flow at low {Reynolds} number.
\newblock {\em Journal of Fluid Mechanics}, 177:133--166, 1987.

\bibitem{kochkov2021}
Dmitrii Kochkov, Jamie~A Smith, Ayya Alieva, Qing Wang, Michael~P Brenner, and Stephan Hoyer.
\newblock Machine learning--accelerated computational fluid dynamics.
\newblock {\em Proceedings of the National Academy of Sciences}, 118(21):e2101784118, 2021.

\bibitem{lagemann2021}
Christian Lagemann, Kai Lagemann, Sach Mukherjee, and Wolfgang Schr{\"o}der.
\newblock Deep recurrent optical flow learning for particle image velocimetry data.
\newblock {\em Nature Machine Intelligence}, 3(7):641--651, 2021.

\bibitem{lagemann2022}
Christian Lagemann, Kai Lagemann, Sach Mukherjee, and Wolfgang Schr{\"o}der.
\newblock Generalization of deep recurrent optical flow estimation for particle-image velocimetry data.
\newblock {\em Measurement Science and Technology}, 33(9):094003, 2022.

\bibitem{lagemann2023learning}
Kai Lagemann, Christian Lagemann, and Sach Mukherjee.
\newblock Learning latent dynamics via invariant decomposition and (spatio-) temporal transformers.
\newblock {\em arXiv preprint arXiv:2306.12077}, 2023.

\bibitem{lagemann2022deep}
Kai Lagemann, Christian Lagemann, Bernd Taschler, and Sach Mukherjee.
\newblock Deep learning of causal structures in high dimensions.
\newblock {\em arXiv preprint arXiv:2212.04866}, 2022.

\bibitem{lecordier2004europiv}
Bertrand Lecordier and Jerry Westerweel.
\newblock The {EUROPIV} synthetic image generator ({S.I.G.}).
\newblock In {\em Particle image velocimetry: recent improvements}, pages 145--161. Springer, Berlin, Heidelberg, 2004.

\bibitem{lee2017}
Hyun~Jung Lee, Miguel~F Diaz, Katherine~M Price, Joyce~A Ozuna, Songlin Zhang, Eva~M Sevick-Muraca, John~P Hagan, and Pamela~L Wenzel.
\newblock Fluid shear stress activates {YAP1} to promote cancer cell motility.
\newblock {\em Nature Communications}, 8(1):14122, 2017.

\bibitem{lee2013petascale}
Myoungkyu Lee, Nicholas Malaya, and Robert~D Moser.
\newblock Petascale direct numerical simulation of turbulent channel flow on up to 786k cores.
\newblock In {\em Proceedings of the International Conference on High Performance Computing, Networking, Storage and Analysis}, pages 1--11, 2013.

\bibitem{li2008}
Yi~Li, Eric Perlman, Minping Wan, Yunke Yang, Charles Meneveau, Randal Burns, Shiyi Chen, Alexander Szalay, and Gregory Eyink.
\newblock A public turbulence database cluster and applications to study lagrangian evolution of velocity increments in turbulence.
\newblock {\em Journal of Turbulence}, 9:N31, 2008.

\bibitem{li2020}
Zongyi Li, Nikola~Borislavov Kovachki, Kamyar Azizzadenesheli, Burigede liu, Kaushik Bhattacharya, Andrew Stuart, and Anima Anandkumar.
\newblock Fourier neural operator for parametric partial differential equations.
\newblock In {\em International Conference on Learning Representations}, 2021.

\bibitem{liou1993}
Meng-Sing Liou and Christopher~J Steffen~Jr.
\newblock A new flux splitting scheme.
\newblock {\em Journal of Computational physics}, 107(1):23--39, 1993.

\bibitem{loshchilov2018fixing}
Ilya Loshchilov and Frank Hutter.
\newblock Decoupled weigth decay regularization.
\newblock In {\em The Seventh International Conference on Learning Representations}, 2019.

\bibitem{lunte2020}
Jens Lunte and Erich Sch{\"u}lein.
\newblock Wall shear stress measurements by white-light oil-film interferometry.
\newblock {\em Experiments in Fluids}, 61:1--12, 2020.

\bibitem{marquardt2019}
Pascal Marquardt, Michael Klaas, and Wolfgang Schr{\"o}der.
\newblock Experimental investigation of isoenergetic film-cooling flows with shock interaction.
\newblock {\em AIAA Journal}, 57(9):3910--3923, 2019.

\bibitem{marquardt2020experimental}
Pascal Marquardt, Michael Klaas, and Wolfgang Schr{\"o}der.
\newblock Experimental investigation of the turbulent {Schmidt} number in supersonic film cooling with shock interaction.
\newblock {\em Experiments in Fluids}, 61(7):160, 2020.

\bibitem{marusic2021}
Ivan Marusic, Dileep Chandran, Amirreza Rouhi, Matt~K Fu, David Wine, Brian Holloway, Daniel Chung, and Alexander~J Smits.
\newblock An energy-efficient pathway to turbulent drag reduction.
\newblock {\em Nature Communications}, 12(1):1--8, 2021.

\bibitem{marusic2010predictive}
Ivan Marusic, Romain Mathis, and Nicholas Hutchins.
\newblock Predictive model for wall-bounded turbulent flow.
\newblock {\em Science}, 329(5988):193--196, 2010.

\bibitem{maeteling2023}
Esther M\"ateling, Marian Albers, and Wolfgang Schr\"oder.
\newblock How spanwise travelling transversal surface waves change the near-wall flow.
\newblock {\em Journal of Fluid Mechanics}, 957:A30, 2023.

\bibitem{maeteling2022}
Esther M\"ateling and Wolfgang Schr\"oder.
\newblock Analysis of spatiotemporal inner-outer large-scale interactions in turbulent channel flow by multivariate empirical mode decomposition.
\newblock {\em Physical Review Fluids}, 7(3):034603, 2022.

\bibitem{mathis2011}
Romain Mathis, Nicholas Hutchins, and Ivan Marusic.
\newblock A predictive inner--outer model for streamwise turbulence statistics in wall-bounded flows.
\newblock {\em Journal of Fluid Mechanics}, 681:537--566, 2011.

\bibitem{mathis2013}
Romain Mathis, Ivan Marusic, Sergei~I Chernyshenko, and Nicholas Hutchins.
\newblock Estimating wall-shear-stress fluctuations given an outer region input.
\newblock {\em Journal of Fluid Mechanics}, 715:163--180, 2013.

\bibitem{meinke2002comparison}
Matthias Meinke, Wolfgang Schr{\"o}der, Ewald Krause, and Th~Rister.
\newblock A comparison of second-and sixth-order methods for large-eddy simulations.
\newblock {\em Computers \& Fluids}, 31(4-7):695--718, 2002.

\bibitem{musa2018}
Mirko Musa, Craig Hill, Fotis Sotiropoulos, and Michele Guala.
\newblock Performance and resilience of hydrokinetic turbine arrays under large migrating fluvial bedforms.
\newblock {\em Nature Energy}, 3(10):839--846, 2018.

\bibitem{orlu2020}
Ramis {\"O}rl{\"u} and Ricardo Vinuesa.
\newblock Instantaneous wall-shear-stress measurements: advances and application to near-wall extreme events.
\newblock {\em Measurement Science and Technology}, 31(11):112001, 2020.

\bibitem{adam_paszke_automatic_2017}
Adam Paszke, Sam Gross, Francisco Massa, Adam Lerer, James Bradbury, Gregory Chanan, Trevor Killeen, Zeming Lin, Natalia Gimelshein, Luca Antiga, Alban Desmaison, Andreas Kopf, Edward Yang, Zachary DeVito, Martin Raison, Alykhan Tejani, Sasank Chilamkurthy, Benoit Steiner, Lu~Fang, Junjie Bai, and Soumith Chintala.
\newblock Pytorch: An imperative style, high-performance deep learning library.
\newblock In {\em Advances in Neural Information Processing Systems}, volume~32, 2019.

\bibitem{perlman2007}
Eric Perlman, Randal Burns, Yi~Li, and Charles Meneveau.
\newblock Data exploration of turbulence simulations using a database cluster.
\newblock In {\em Proceedings of the 2007 ACM/IEEE Conference on Supercomputing}, pages 1--11, 2007.

\bibitem{pope2000}
Stephen~B Pope.
\newblock {\em Turbulent flows}.
\newblock Cambridge University Press, Cambridge, 2000.

\bibitem{quadrio2022}
Maurizio Quadrio, Alessandro Chiarini, Jacopo Banchetti, Davide Gatti, Antonio Memmolo, and Sergio Pirozzoli.
\newblock Drag reduction on a transonic airfoil.
\newblock {\em Journal of Fluid Mechanics}, 942:R2, 2022.

\bibitem{raffel_particle_2018}
Markus Raffel, Christian~E Willert, Fulvio Scarano, Christian~J K{\"a}hler, Steve~T Wereley, and J{\"u}rgen Kompenhans.
\newblock {\em Particle Image Velocimetry: A Practical Guide}.
\newblock Springer International Publishing AG, Springer Nature, 2018.

\bibitem{raissi2019}
Maziar Raissi, Paris Perdikaris, and George~E Karniadakis.
\newblock Physics-informed neural networks: A deep learning framework for solving forward and inverse problems involving nonlinear partial differential equations.
\newblock {\em Journal of Computational physics}, 378:686--707, 2019.

\bibitem{Ricco2021}
Pierre Ricco, Martin Skote, and Michael Leschziner.
\newblock A review of turbulent skin-friction drag reduction by near-wall transverse forcing.
\newblock {\em Progress in Aerospace Sciences}, 123:100713, 2021.

\bibitem{rubbert2019}
Alexander Rubbert, Marian Albers, and Wolfgang Schr{\"o}der.
\newblock Streamline segment statistics propagation in inhomogeneous turbulence.
\newblock {\em Physical Review Fluids}, 4(3):034605, 2019.

\bibitem{schaefer2011}
Lisa Sch\"afer, Uwe Dierksheide, Michael Klaas, and Wolfgang Schr\"oder.
\newblock Investigation of dissipation elements in a fully developed turbulent channel flow by tomographic particle-image velocimetry.
\newblock {\em Physics of Fluids}, 23(3):035106, 2011.

\bibitem{scharnowski2020}
Sven Scharnowski and Christian~J K{\"a}hler.
\newblock Particle image velocimetry-classical operating rules from today’s perspective.
\newblock {\em Optics and Lasers in Engineering}, 135:106185, 2020.

\bibitem{schrijer_effect_2008}
Ferry Schrijer and Fulvio Scarano.
\newblock Effect of predictor--corrector filtering on the stability and spatial resolution of iterative {PIV} interrogation.
\newblock {\em Experiments in Fluids}, 45(5):927--941, 2008.

\bibitem{segalini2015}
Antonio Segalini, Jean-Daniel R{\"u}edi, and Peter~A Monkewitz.
\newblock Systematic errors of skin-friction measurements by oil-film interferometry.
\newblock {\em Journal of Fluid Mechanics}, 773:298--326, 2015.

\bibitem{siam2017convolutional}
Mennatullah Siam, Sepehr Valipour, Martin Jagersand, and Nilanjan Ray.
\newblock Convolutional gated recurrent networks for video segmentation.
\newblock In {\em 2017 IEEE International Conference on Image Processing (ICIP)}, pages 3090--3094. IEEE, 2017.

\bibitem{souilhol2020}
Celine Souilhol, Jovana Serbanovic-Canic, Maria Fragiadaki, Timothy~J Chico, Victoria Ridger, Hannah Roddie, and Paul~C Evans.
\newblock Endothelial responses to shear stress in atherosclerosis: a novel role for developmental genes.
\newblock {\em Nature Reviews Cardiology}, 17(1):52--63, 2020.

\bibitem{sparreboom2009principles}
Wouter Sparreboom, Albert van~den Berg, and Jan~CT Eijkel.
\newblock Principles and applications of nanofluidic transport.
\newblock {\em Nature Nanotechnology}, 4(11):713--720, 2009.

\bibitem{sturzebecher2001}
D~Sturzebecher, S~Anders, and W~Nitsche.
\newblock The surface hot wire as a means of measuring mean and fluctuating wall shear stress.
\newblock {\em Experiments in Fluids}, 31(3):294--301, 2001.

\bibitem{tanner1976}
LH~Tanner and LG~Blows.
\newblock A study of the motion of oil films on surfaces in air flow, with application to the measurement of skin friction.
\newblock {\em Journal of Physics E: Scientific Instruments}, 9(3):194, 1976.

\bibitem{teed2020}
Zachary Teed and Jia Deng.
\newblock Raft: Recurrent all-pairs field transforms for optical flow.
\newblock In {\em Computer Vision--ECCV 2020: 16th European Conference}, pages 402--419. Springer, 2020.

\bibitem{tzima2005mechanosensory}
Eleni Tzima, Mohamed Irani-Tehrani, William~B Kiosses, Elizabetta Dejana, David~A Schultz, Britta Engelhardt, Gaoyuan Cao, Horace DeLisser, and Martin~Alexander Schwartz.
\newblock A mechanosensory complex that mediates the endothelial cell response to fluid shear stress.
\newblock {\em Nature}, 437(7057):426--431, 2005.

\bibitem{vaswani2017attention}
Ashish Vaswani, Noam Shazeer, Niki Parmar, Jakob Uszkoreit, Llion Jones, Aidan~N Gomez, {\L}ukasz Kaiser, and Illia Polosukhin.
\newblock Attention is all you need.
\newblock {\em Advances in Neural Information Processing Systems}, 30, 2017.

\bibitem{vinuesa2022}
Ricardo Vinuesa and Steven~L Brunton.
\newblock Enhancing computational fluid dynamics with machine learning.
\newblock {\em Nature Computational Science}, 2(6):358--366, 2022.

\bibitem{vinuesa2023}
Ricardo Vinuesa, Steven~L Brunton, and Beverley~J McKeon.
\newblock The transformative potential of machine learning for experiments in fluid mechanics.
\newblock {\em Nature Reviews Physics}, 5(9):1--10, 2023.

\bibitem{westerweel2004}
Jerry Westerweel, Peter Geelhoed, and Ralph Lindken.
\newblock Single-pixel resolution ensemble correlation for micro-{PIV} applications.
\newblock {\em Experiments in Fluids}, 37(3):375--384, 2004.

\bibitem{womersley1955}
John~R Womersley.
\newblock Method for the calculation of velocity, rate of flow and viscous drag in arteries when the pressure gradient is known.
\newblock {\em The Journal of Physiology}, 127(3):553, 1955.

\bibitem{yang2013}
Chi Yang, Fuxin Huang, and Francis Noblesse.
\newblock Practical evaluation of the drag of a ship for design and optimization.
\newblock {\em Journal of Hydrodynamics}, 25(5):645--654, 2013.

\bibitem{zhou2017}
Geng Zhou, Yueqi Zhu, Yanling Yin, Ming Su, and Minghua Li.
\newblock Association of wall shear stress with intracranial aneurysm rupture: systematic review and meta-analysis.
\newblock {\em Scientific Reports}, 7(1):1--8, 2017.

\end{thebibliography}

\newpage
\section*{Supplementary Information}
Here, we provide details about the datasets used in this work, i.e., the measurement setups and relevant aspects of the numerical simulations and analytic solutions. Moreover, we detail on the rendering pipeline of the synthetic particle images.

\subsection*{PIV measurement data}
\begin{wrapfigure}{r}{0.62\textwidth}
  \vspace{-0.6cm}
  \begin{center}
   \includegraphics[width=.63\textwidth, trim={0.15cm 0.35cm 0.15cm  0.35cm},clip]{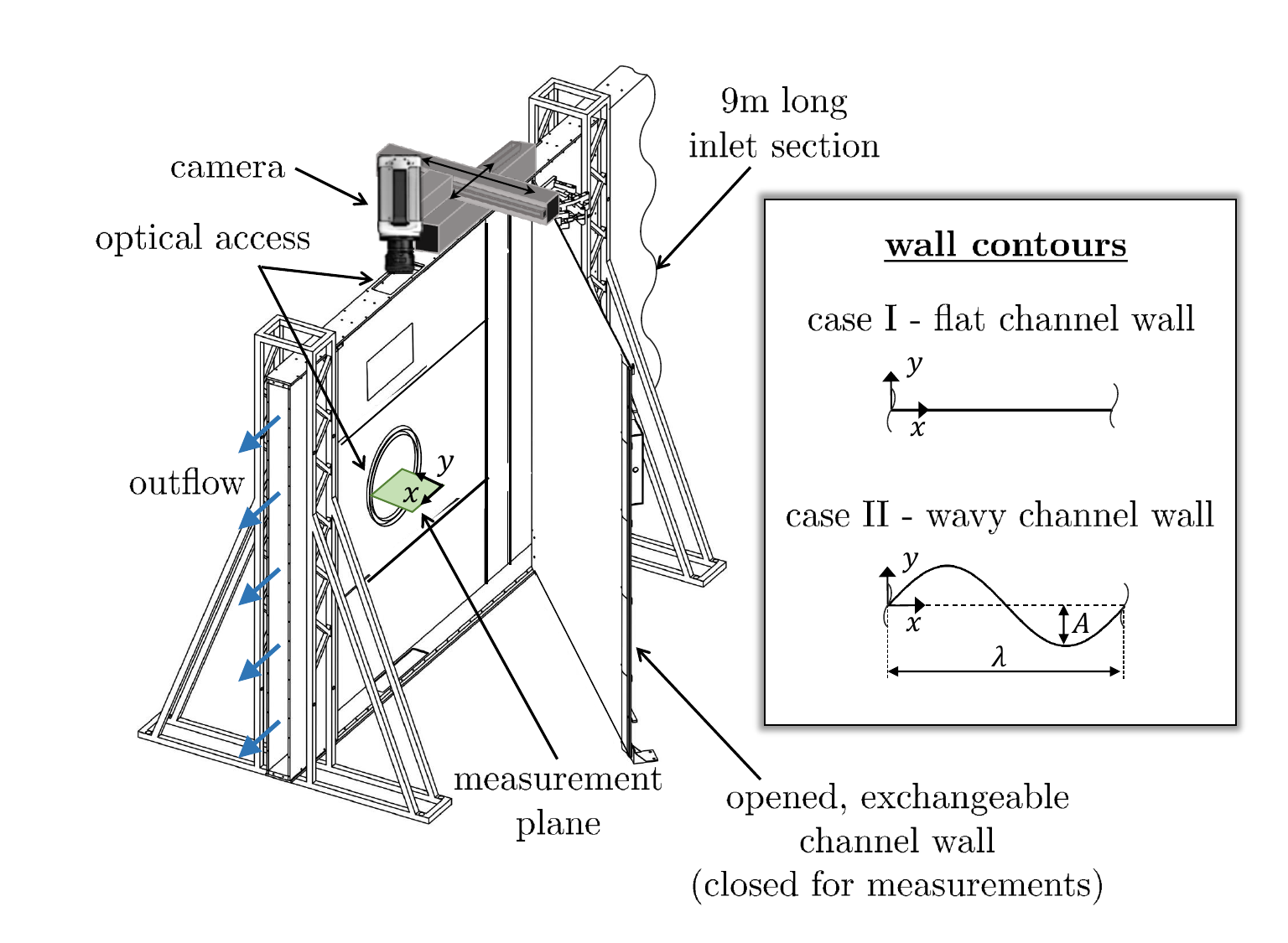}
  \end{center}
	 \caption{Measurement section of the TCF facility.}
        \label{fig:setup}
\end{wrapfigure}

\textbf{Turbulent channel and turbulent wavy channel flow.} 
The experiments are performed in an Eiffel-type wind tunnel at the Institute of Aerodynamics at RWTH Aachen University. A $9000$~mm long inlet section with a tripping device provides a fully developed turbulent channel flow (TCF) in the $2700$~mm long measurement section, which is shown in figure~\ref{fig:setup}. It features an aspect ratio of $AR$~$=$~$20$ with a cross-section of $100$~x~$2000$~mm$^2$ (height $2h$ x width $w$) that ensures a negligible influence of three-dimensional effects. Comparisons to DNS data \citep{schaefer2011} verify the quality of the provided fully developed TCF. \\
A key feature of this wind tunnel is its exchangeable side wall. In addition to the flat wall TCF measurements (case I in figure~\ref{fig:setup}), investigations with a sinusoidally shaped side wall (case II in figure~\ref{fig:setup}) are conducted. This wavy TCF exhibits changing pressure gradients along the streamwise direction creating a generic yet challenging flow topology. In the present study, the sine wave features a wavelength of $\lambda = 2h = 100$~mm and an amplitude of $A = 0.1h = 5$~mm. The trough of the wave is aligned with the flat channel wall position such that the crests of the wavy wall are located at $y = 10$~mm. Similar measurements have already been conducted by \cite{rubbert2019}.\\ 
For the flat TCF measurements, the two-dimensional two-component PIV setup is realized with a \textit{Darwin Duo 40} laser and a \textit{Photron SA3} camera equipped with a \textit{Tamron 180~mm} lens. The input image size of this test case is $1024 \times 1024 \, \mathrm{px}^2$ and the camera operates at a frame rate of $1000~\mathrm{Hz}$. The friction Reynolds number is $Re_{\tau} \approx 1120$ and the viscous sublayer is captured by $3$~px. 

The flow in the wavy TCF is recorded using a \textit{New Wave Solo 200XT} laser and a \textit{PCO edge 5.5 sCMOS} camera equipped with a \textit{Zeiss 100 mm} macro lens. This setup results in an input image size of $2160 \times 2560~\mathrm{px}^2$ with a very high spatial resolution of the flow field such that the viscous sublayer is covered by a minimum of $14$~px at $Re_{\tau} \approx 200$. The acquisition frequency of the wavy TCF test case is $15$~Hz. In both setups, the measurement plane is oriented in the streamwise ($x$) and wall-normal ($y$) direction at the channel's centerline as shown in figure~\ref{fig:setup}. In total, $2000$ image pairs are recorded and evaluated.\\

\begin{figure}[h]
    \centering
    \includegraphics[width=0.7\textwidth]{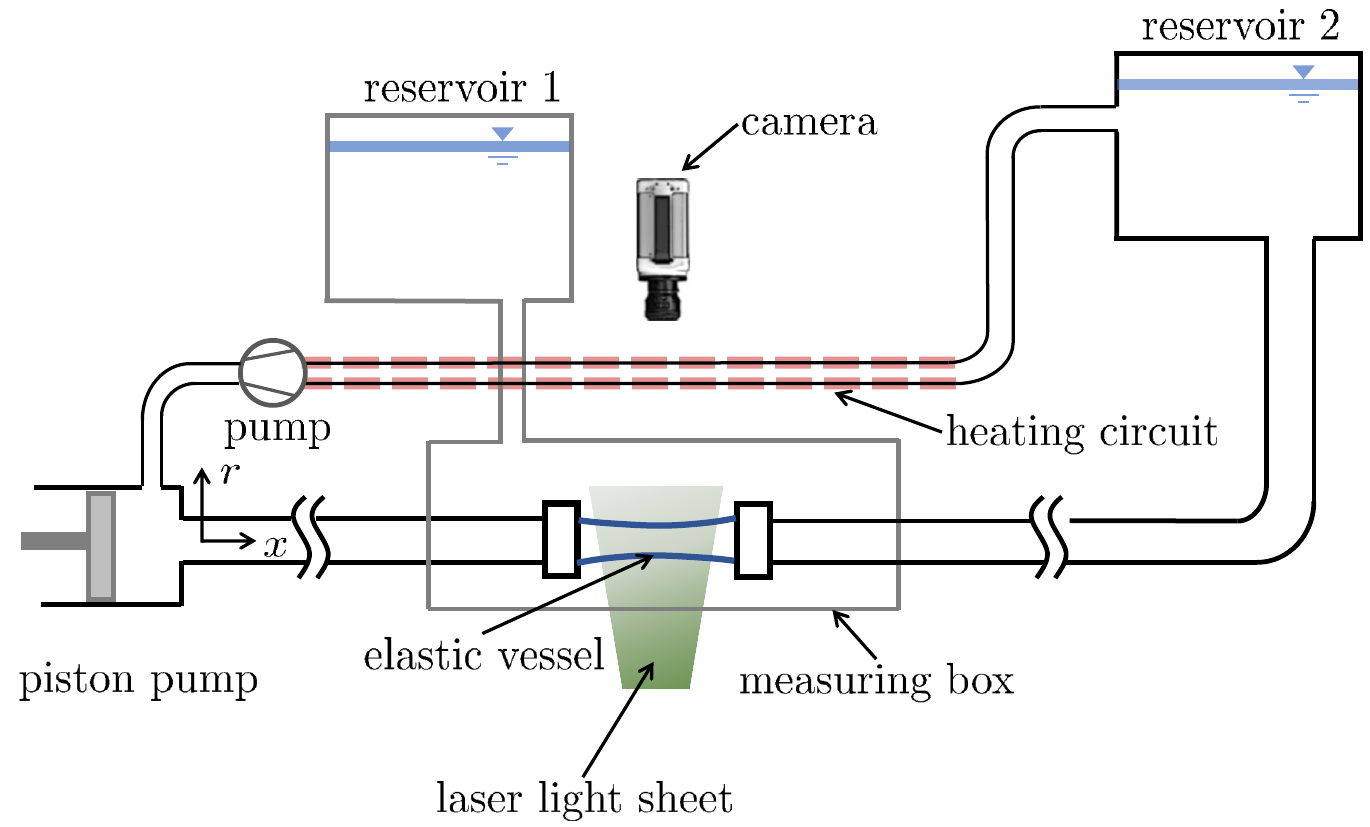}
    \caption{Experimental setup of the blood vessel model.}
    \label{fig:setup_vessel}
\end{figure}

\textbf{Elastic blood vessel flow.}
As shown in the sketch of the experimental setup (figure~\ref{fig:setup_vessel}), an oscillatory flow is generated in the closed circuit by a controlled piston pump. The fluid flow is a water-glycerin mixture, which approximates the blood flow properties and matches the refractive index of the vessel model such that no optical distortions bias the PIV measurements. The elastic vessel model is made from \textit{RTV615}, which is a Polydimethylsiloxane (PDMS) silicone polymer. It is mounted in a rectangular container which is filled with the same water-glycerin mixture as the oscillatory circuit to adjust the transmural pressure, i.e., the static pressure difference across the vessel wall. More precisely, the height difference between two reservoirs (see figure~\ref{fig:setup_vessel}), one connected to the vessel box (reservoir 1) and one to the flow circuit (reservoir 2), is used to adapt the transmural pressure $p_{trans} \approx 12 467~\mathrm{Pa}$. Upstream and downstream of the vessel model, rigid Polymethylmethacrylate (PMMA) pipes with a sufficient length-over-diameter ratio $L/D \approx 138$ ensure a fully developed flow. Since the fluid properties, e.g., the viscosity and the refractive index, are temperature dependent, the flow circuit is equipped with a temperature control system. Prior to the measurements, the fluid is heated up to $25.6^\circ \mathrm{C}$ and a well-designed insulation ensures a negligible heat loss throughout the circuit. The measuring box is heated separately using heating films while a small pump inside the box ensures a homogeneous temperature distribution. \\

The PIV measurements are conducted in the center of the vessel model with the measurement plane oriented in the streamwise ($x$) and radial ($r$) direction. Hollow glass spheres with a mean diameter of $D \approx 10~\mu \mathrm{m}$ and a density of $\rho =1100~ \mathrm{kg/m^3}$ are used as tracer particles. A \textit{Darwin Duo 40} laser and a \textit{Fastcam PCI 1024} camera equipped with a \textit{Tamron 180~mm} lens are synchronized with the phase angle of the oscillatory flow via the position sensoring system of the linear motor of the piston pump. The particle image input size is $1024 \times 1024 ~\mathrm{px}^2$ and the frame rate measures $200$~Hz. \\
The vessel properties and the flow conditions are chosen to approximate the behavior of the human abdominal aorta. That is, the Reynolds number based on the vessel diameter $D_v$ and the maximum fluid velocity $\bar{u}_{max}$ measures $Re_v = \rho \bar{u}_{max} D_v / \eta \approx 390$ and the Womersley number based on the vessel radius $R_v = D_v/2$ and the angular frequency of the piston pump $f_p$ is $Wo = R_v \sqrt{\rho 2 \pi f_p / \eta} \approx 5.27$. \\

\subsection*{Baseline data}
\textbf{Turbulent channel flow.}
The DNS data used for validating the experimental turbulent channel flow results and generating the synthetic particle images are provided by the Johns Hopkins Turbulence Databases \cite{li2008,perlman2007}. In the following, computational details of this simulation are provided but we refer the interested reader to the original publications of \cite{li2008,perlman2007,graham2016} for a comprehensive data description. \\
The DNS data contain a turbulent channel flow at $Re_\tau \approx 1000$ within a computational domain of $L_x \times L_y \times L_z = 8\pi h \times 2 h \times 3\pi h$, where $h$ denotes the channel half-height and $x,y,z$ the streamwise, wall-normal, and spanwise direction. The numerical grid contains $N_x \times N_y \times N_z = 2048 \times 512 \times 1536$ grid points and the numerical time step is $\Delta t = 0.0013$ in dimensionless units. Periodic boundary conditions are applied in the longitudinal and transverse directions and no-slip conditions at the top and bottom walls. The Navier-Stokes equations are solved using a wall–normal velocity–vorticity formulation \cite{kim1987turbulence}. Solutions to the governing equations are obtained using a Fourier-Galerkin pseudo-spectral method for the longitudinal and transverse directions and a seventh-order Basis-splines collocation method in the wall-normal direction. Dealiasing is performed via the 3/2-rule \cite{lee2013petascale}. For the temporal integration, a low-storage, third-order Runge-Kutta method is used. Initially, the flow is driven via
a constant volume flux control, i.e., a bulk mean velocity of $\bar{u} = 1$ is imposed until stationary conditions are reached. Then, the control is changed to a prescribed mean pressure gradient. Additional iterations are performed to further achieve statistical stationarity before outputting the flow fields. Overall, the available simulation data cover approximately a single flow through time with the velocity and pressure fields stored every five time steps resulting in $4000$ frames of data.\\

\textbf{Turbulent wavy channel flow.}
Numerical data of the wavy turbulent channel flow are obtained by a DNS using the high-performance in-house flow solver m-AIA (multiphysics-Aerodynamisches Institut Aachen). The dimensions of the computational domain are $L_x \times L_y \times L_z = 8 h \times 2 h \times 4 h$, where $h$ denotes the channel half-height and $x,y,z$ the streamwise, wall-normal, and spanwise direction. The wavy surface of the lower channel wall has a wavelength of $\lambda = 2h = 100 \mathrm{mm}$ such that four waves are included within the streamwise direction. Directly at the wall, a grid resolution of $\Delta x^+ = 4.1$, $\Delta y^+ = 0.5$ and $\Delta z^+ = 8.2$ is deployed, whereas the wall-normal resolution is coarsen with increasing wall distance. This yields a numerical grid with $N_x \times N_y \times N_z = 400 \times 180 \times 100 = 4.8 \times 10^6$ cells. Periodic boundary conditions are applied in the longitudinal and transverse direction and no-slip boundary conditions at the flat and wavy sidewalls. The friction Reynolds number is approximately $Re_\tau \approx 200$ to match the experimental conditions. \\
The compressible unsteady Navier-Stokes equations are solved directly (DNS) using the monotonically integrated large-eddy simulation (MILES) approach~\citep{boris1992,fureby1999}. The intrinsic numerical dissipation of the inviscid flux discretization, which is computed by the advection upstream splitting method (AUSM)~\citep{liou1993}, is sufficient to provide the required turbulent dissipation of the smallest unresolved scales. The temporal integration is performed by an explicit five-stage Runge-Kutta method of second-order accuracy. Among numerous other studies, investigations by \cite{meinke2002comparison} have confirmed the high accuracy of the flow solver m-AIA. Further details of the wavy turbulent channel flow simulations can be found in \cite{rubbert2019}.\\

\textbf{Womersley solution of a blood vessel flow.}
The so-called Womersley solution refers to a publication by Womersley~\citep{womersley1955} in which he derived a formula to calculate the oscillatory arterial flow. The formula constitutes an exact solution of the equations of viscous fluid motion in a circular pipe experiencing a periodic pressure gradient, i.e., an oscillatory flow. The resulting streamwise velocity distribution is calculated via
\begin{equation}
    u(r,\phi) = \mathcal{R} \left( \frac{A_p}{i \omega \rho} \left[ 1 - \frac{J_0 \left( Wo \cdot r/R_v \cdot i^{3/2} \right)}{J_0 \left( Wo \cdot i^{3/2} \right)} \right] e^{i \phi} \right),
\end{equation}
where $\mathcal{R}$ denotes the real part of the solution. The periodic pressure gradient is represented by $A_p e^{i \phi}$, where $A_p = |dp/dx|$ defines the amplitude via the streamwise pressure gradient and $e^{i \phi}$ is the periodic distribution. The quantity $\omega = 2 \pi f_s$ refers to the angular frequency of the oscillation, which is calculated with the frequency $f_s = 0.2$~Hz of the piston pump to match the experimental conditions. The function $J_0$ represents the bessel function of first kind and order zero. The Womersley number $Wo = R_v \sqrt{\frac{\rho \omega}{\eta}}$ is calculated with the radius of the vessel $R_v = 11.14$~mm and the fluid properties, i.e., the density $\rho = 1143.3$~kg/m$^3$ and the dynamic viscosity $\eta = 6.422$~mPas, which represent the conditions of the water-glycerin mixture at $T \approx 25.6^\circ C$. This temperature is chosen to approximate the blood flow properties at $T \approx 37^\circ C$ in the experimental setting.

\subsection*{Rendering pipeline of synthetic particle images}
In this work, a novel training dataset is generated that targets to represent realistic conditions with respect to the comprised flow fields, tracer particle appearances, and lighting conditions. Thus, it aims at a broader generalization ability to arbitrary real-world PIV experiments than previously introduced synthetic particle images by, e.g., \cite{cai2019,lagemann2021}. Ground-truth velocity fields are taken from in-house conducted and publicly available DNS data. The flow fields cover a series of test cases that are important to a large number of academic, physics, and engineering related PIV applications, e.g., homogeneous isotropic turbulence, magneto-hydrodynamic isotropic turbulence, homogeneous buoyancy driven turbulence, laminar and turbulent boundary layer flows, and turbulent channel flows at various Reynolds numbers. \\
An optimized particle image shifting and image rendering process is used to achieve more realistic particle images. The synthetic images are generated with a similar procedure as described in \cite{lecordier2004europiv}. To ensure that the generated particle images move along physical streamlines, a fourth-order Runge-Kutta integration scheme is applied when shifting the particle image locations within the image pair. According to \cite{raffel_particle_2018}, the grey value intensity distribution of the particle images can be assumed to be Gaussian distributed, where $\sigma$ denotes the standard deviation of the particle image diameter. The following integral approach for the particle image intensity is chosen

\begin{equation}
\resizebox{.9\hsize}{!}{$I(x,y)= I_p \pi \sigma^2 \frac{\left [erf \left (\frac{(x_p-x_{min})}{(\sqrt2 \sigma)} \right )-erf \left (\frac{(x_p-x_{max})}{(\sqrt2 \sigma)} \right )\right ] \cdot \left [erf \left (\frac{(y_p-y_{min})}{(\sqrt2 \sigma)} \right )-erf \left (\frac{(y_p-y_{max})}{(\sqrt2 \sigma)} \right )\right ] }{2} $}    
\end{equation}
with the error function
\begin{equation}
    erf(\textbf{a})=\frac{2}{\sqrt{\pi}}\int_{0}^{\textbf{a}}e^{t^2}dt,
\end{equation}

where $I_p$ is the particle image intensity and $(x_p, y_p)$ are the coordinates of each pixel center within the integration domain. The quantities $x_{min}$, $x_{max}$, $y_{min}, y_{max}$ denote the integration boundaries  of each pixel located at $[x_p,y_p] \pm 0.5 \,\mathrm{px}$.\\

To generate a real-world inspired and realistic synthetic training dataset, properties of characteristic particle image and lighting conditions are extracted from experimentally obtained PIV data. That is, numerous in-house conducted and open-source available PIV image datasets are thoroughly studied and relevant parameters, e.g., the particle image diameter range and the corresponding particle image shape, the seeding density, and the signal-to-noise ratio (SNR) including background noise intensities, are extracted. Subsequently, image rendering parameters are sampled from the obtained parameter ranges and as a result, realistic synthetic PIV images are created. To summarize, the particle image diameters range from $0.2 \, \mathrm{px} < D_p < 5 \, \mathrm{px}$ while the particle image density $N_{ppp}$, which equals the number of particle images per pixel, varies in the range of $0.005 < N_{ppp} < 0.1$. The SNR is sampled in the range of $1 < SNR < 16$. The maximum particle image displacement is set to $\pm 16\,\mathrm{px}$ and Gaussian image noise is added with a mean $\mu_n=0.01$ and a variance $var_n=0.075$ based on the normalized pixel values of each individual image. Overall, approximately $200,000$ individual double frames are rendered for training and $57,000$ image pairs for validation containing $256 \times 256 \,\mathrm{px}^2$. In summary, this very large training dataset mimics real-world PIV images realistically with a depth and variation going far beyond currently available datasets. \\
The synthetic particle images of the flat and the wavy turbulent channel flows are generated using the same rendering approach as described above. The underlying ground-truth velocity distributions are taken from DNS data that closely match the experimental measurement conditions to provide meaningful baseline data for comparison.

\end{document}